# Seasonal Evolution of Mixed Layers in the Red Sea and the Relative Contribution of Atmospheric Buoyancy and Momentum Forcing

G. Krokos[1], I. Cerovečki[2], P. Zhan[1], M. C. Hendershott[2] and I. Hoteit[1]

[1]King Abdullah University of Science and Technology, Thuwal, Saudi Arabia.

[2]Scripps Institution of Oceanography, University of California, San Diego, USA.

Corresponding author: Ibrahim Hoteit (ibrahim.hoteit@kaust.edu.sa)

**Key Points:**

- Mixed layers in the Red Sea and their seasonal evolution are examined via a high-resolution ocean circulation model
- The seasonal evolution of Red Sea mixed layer depths is primarily driven by the air-sea buoyancy forcing, especially its heat flux component
- Surface winds intensified by topographic features strongly influence the local mixed layer depths

**Abstract**

The seasonal and spatial evolution of mixed layers (MLs) in the Red Sea (RS) is analyzed for the 2001–2015 period using the results of a high resolution (1/100°, 50 vertical layers) ocean circulation model forced by a novel regional high resolution (5 km) atmospheric reanalysis dataset. The simulation reproduces the main features of the near-surface stratification, as described by the available observations. The seasonal evolution of the modeled mixed layer depths (MLDs) in the RS is predominantly driven by atmospheric buoyancy forcing, especially its heat flux component. Everywhere in the basin the model MLs are deepest in January and February. The deepest MLDs develop in the northern parts of the Gulf of Aqaba and in the western parts of the north RS. The MLDs gradually shoal towards the south, reflecting the meridional gradient of wintertime surface buoyancy loss. In spring and summer, the surface ocean heat gain increases the stratification and the MLs are becoming shallow everywhere in the basin. During this season wind may have a significant local impact on the MLD. Particularly important are strong winds channeled by topography, such as in the vicinity of the Strait of Bab-Al-Mandeb and the straits connecting the two gulfs in the north, and lateral jets blowing through mountain gaps, such as the Tokar jet in the central RS. The MLD distribution further suggests influence by the general and mesoscale circulation. The complex patterns of air–sea buoyancy flux, wind forcing, and the thermohaline and mesoscale circulation, are all strongly imprinted on the MLD distribution.

**Plain Language Summary**

The air–sea exchange of heat, water, and gases, such as oxygen and carbon dioxide, occurs through the near-surface vertically well-mixed layer. Mixing in the upper layer redistributes its properties and mediates the sequestration of heat and gases from the atmosphere into the sea, influencing its physical, chemical, and biological characteristics. In this study, using a high-resolution ocean circulation model, we investigated the seasonal evolution of the Red Sea (RS) mixed layer depths (MLDs). Our results show strong spatial and seasonal MLD variability, with the deepest MLDs developing in the northern RS, as a result of intense wintertime cooling of highly saline waters. In spring and summer, the surface ocean heat gain increases the stratification and MLs are shallow everywhere in the basin. During that time, deeper MLs develop locally driven by wind. Particularly important are strong winds channeled by topography. The shallowest MLDs are found in the southern RS, where their seasonal evolution is influenced by the exchange of water through the Strait of Bab-Al-Mandeb. The complex patterns of atmospheric forcing and the general and mesoscale circulation are all imprinted on the MLD distribution.

## 1. Introduction

Located between Africa and the Arabian Peninsula, the Red Sea (RS) is an elongated semi-enclosed basin that extends from 12.5°N to 30°N in the SE–NW direction, with an average width of 220 km (Patzert, 1974). Although the maximum depths along the axial trench exceed 3000 m, very shallow (less than 100 m) and broad shelf platforms cover more than 40% of the basin, especially in the Southern RS (SRS) (Rasul et al., 2015). The Gulfs of Aqaba (hereafter GoAq) and Suez (hereafter GoS) are located at the northeast and northwest tips of the North RS (NRS), respectively (Fig. 1). The GoS is very shallow, with an average depth of 40 m, and a mean width of approximately 30 km; it extends to the northwest for approximately 300 km, and is connected to the RS via a wide shelf. By contrast, depths in the GoAq exceed 1,800 m (Neumann and McGill, 1962). This Gulf is connected to the RS through the narrow Straits of Tiran, with a sill depth of approximately 260 m. In the southern RS, a relatively shallow (maximum 137 m depth) and narrow (18 km) constriction, the Strait of Bab-Al-Mandeb (BAM), connects the RS to the Gulf of Aden (GoAd) and the Indian Ocean (Fig. 1).

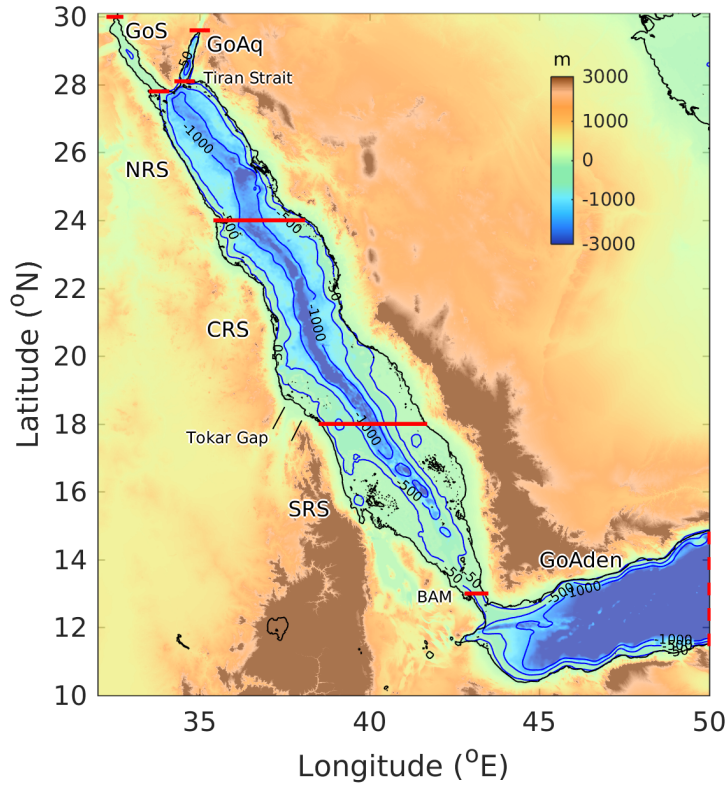

*Figure 1. Bathymetric map of the model domain, covering the Red Sea (RS) and the adjacent Gulfs. Red lines indicate the boundaries of the individual analysis regions: NRS, CRS, SRS, GoS and GoAd. The dashed red line indicates the model open boundary.*

The general circulation in the RS is mainly driven by air–sea buoyancy fluxes and winds (e.g., Yao et al., 2014a,b; Sofianos and Johns, 2015). Although the RS is situated in the tropics, between two of the world's largest deserts, it exhibits a weak annual net heat loss (7 W/m$^2$, Patzert, 1974; 11 ± 5 W/m$^2$, Sofianos et al., 2002). Because of its arid environment, the RS experiences one of the highest evaporation rates among the world's oceans, with an annual average of approximately 2 m/year, while precipitation is negligible (Privett, 1959; Morcos, 1982; Tragou et al., 1999; Sofianos et al., 2002; Albarakati and Ahmad, 2013; Ahmad and Albarakati, 2015).

The strongest air–sea buoyancy exchange occurs in winter in the northern RS, driving convective mixing and the formation of two highly saline water masses: the relatively lighter RS intermediate water (RSIW) that can be found below the surface waters at depths up to about 250 m, and the denser RS deep water (RSDW) below it (Phillips, 1966; Cember, 1988; Sofianos and Johns, 2003; Yao et al., 2014a,b). These two water masses gradually flow southwards, with the RSIW constituting most of the dense outflow through the BAM (see also Fig. 2 in Sofianos and Johns, 2015). The outflow of dense waters from the RS through the BAM is balanced by the inflow of fresher waters from the GoAd (Sofianos and Johns, 2007). The exchanges at the BAM exhibit a distinct seasonal pattern governed by the monsoon wind reversal and the associated upwelling in the GoAd (Patzert, 1974). During the winter period, the water exchanges exhibit a two-layer pattern, whereby a deep outflow is balanced by the inflow of surface waters from the GoAd. During the summer period, the mean surface circulation reverses direction towards the GoAd (Yao et al., 2014a,b). The balance is maintained by intrusion of the fresher and cooler GoAd intermediate water (GAIW) at intermediate depths (see also Fig. S3 in Supplementary Material). A relatively weak deep outflow persists throughout the summer period (Yao et al., 2014a; Xie et al., 2018).

The upper part of the thermohaline circulation in the RS comprise a general northward transport that is initially intensified along the western boundary up to the central RS (CRS), before it transitions to the eastern shore of the RS and continues up to the NRS (Quadfasel and Baudner, 1993; Sofianos, 2002; Yao et al., 2014b; Zhai et al., 2015). The northward upper layer circulation intensifies in winter when surface waters enter the RS through the BAM. The upper layer waters are partially transformed in the NRS, and either return southward through deeper layers or intrude further into the GoAq and GoS, balanced by deeper outflows through the straits of the two gulfs.

Mesoscale eddies are present throughout the RS, being particularly frequent in its central region between 18ºN and 24ºN, where they strengthen in winter (Zhan et al., 2014). In the NRS, a permanent cyclonic gyre is centered between 26ºN and 27ºN and it intensifies during winter (Sofianos, 2002; Yao et al., 2014a). This gyre preconditions this region for the intermediate and deep-water formations (Papadopoulos et al., 2015). Cyclone/anticyclone dipoles develop in response to lateral wind jets blowing from mountain gaps surrounding the RS basin, becoming stronger during summer and spring (Jiang et al., 2009; Langodan et al., 2017; Menezes et al., 2018). A quasi-stationary dipolar eddy formed in the CRS in response to the Tokar Jet winds is the most energetic (Bower et al., 2013; Zhai and Bower, 2013; Zhan et al., 2018).

The importance of understanding the RS circulation and its physical properties has been increasingly recognized in recent years owing to the sea's ecological and economical significance, which stems from its remarkable biological diversity and productivity (Carvalho et al., 2019). Of particular importance for the RS ecology is the wintertime surface heat loss in the NRS that drives the mixed layer (ML) deepening and formation of the intermediate and deep waters that ventilate the deep parts of the RS. Deepening of the MLs can entrain nutrient rich waters that enhances the RS productivity (Acker et al., 2008; Gittings et al., 2018, 2019). The ML also plays a crucial role in air–sea exchange of gases such as oxygen, thus impacting the marine life (Gaube et al., 2018). The goal of this work is to provide a comprehensive description of the seasonal and spatial evolution of MLs in the RS.

The ML is characterized by vertically nearly homogeneous water properties such as temperature, salinity and density (Kara, 2003). The air–sea buoyancy loss (given by the sum of the heat and freshwater fluxes) destabilizes stratification and drives convective mixing (Alexander et al., 2000; Kantha and Clayson, 2002; Carton et al., 2008). By contrast, buoyancy gain increases the stability of the surface layers inhibiting ML deepening. The upper layer stratification and the ML distribution are also affected by the general and mesoscale circulation advecting heat and salt (Alexander et al., 2000; Taylor & Ferrari, 2010). Mesoscale eddies tend to homogenize the upper layers by inducing lateral and vertical mixing (Cerovečki and Marshall, 2008; Gaube et al., 2018; Zhan et al., 2019). Eddies can also influence the ML by displacing the isopycnals; cyclonic eddies tend to cause the isopycnals to dome upward, locally decreasing the MLD, whereas anticyclonic eddies tend to deepen the isopycnals, locally increasing the MLD (Dong et al., 2014; Gaube et al., 2018).

While observational studies have provided a general overview of the ML seasonal variability in the RS, due to the paucity of in situ observations, knowledge about MLDs in the RS is still fairly limited (Eladawy et al., 2017). The available observations are very inhomogeneous both in space and time; they have mostly been collected during periods of favorable sea conditions, thus providing a biased representation of the MLD distribution and hindering the identification of local extremes. Abdulla et al. (2018) derived the monthly MLD climatology in the RS based on a historical collection of temperature profiles from in situ observations between 1934 and 2017, suggesting a large spatial and seasonal MLD variability. Particularly deep MLs have been observed in the NRS due to winter cooling of the highly saline surface waters. Wind stress dominates the MLD variability in the CRS and SRS (Abdulla et al. 2018). Ocean eddies and local wind jets, such as that at the Tokar Gap, have also been shown to significantly alter the MLD structure throughout the RS (Zhai and Bower, 2013; Abdulla et al., 2018). In the GoAq, the MLD variability is predominantly driven by variability in air–sea heat exchange (Carlson et al., 2014; Silverman and Gildor, 2008; Wolf-Vecht et al., 1992). Using observations Carlson et al. (2014) showed that MLDs in the GoAq vary from less than 50 m in September to more than 350 m in May.

In this study, we used the results of a regional RS circulation model, forced by a novel downscaled atmospheric product derived from ERA-Interim reanalysis (Dee et al., 2011), for the period of 2001-2015 to study the spatiotemporal variability of the MLs in the RS. The paper is organized as follows: The model setup, the available in situ observations, and the method used to estimate the MLDs are described in Section 2. Section 3 presents the validation of the numerical model results against available observations; additional detailed verification of the model outputs are also provided in the Supplementary material. The seasonal and spatial distributions of wind stress and surface buoyancy fluxes estimated from the model simulation are presented in Section 4; in this section we also qualitatively examine the relative contribution of the wind and surface buoyancy fluxes in inducing vertical mixing, which drives the ML deepening. The seasonal simulated MLD distribution is described in Section 5. Finally, a summary of the main findings concludes the work in Section 6.

## 2. Data and methods

### 2.1. Model configuration

We performed a regional RS hydrodynamic simulation using the Massachusetts Institute of Technology ocean general circulation model (MITgcm) described in Marshall et al., 1997. The model setup is based on that of Yao et al. (2014a,b), but at higher horizontal and vertical resolutions and forced with a higher-resolution atmospheric fields that were specifically generated for the RS region (Hoteit et al., 2021).

The model domain includes the RS, the GoS, the GoAq, and a large part of the GoAd (Fig. 1). The bathymetry dataset is derived from the GEBCO 2014 product (Weatherall et al. 2015). The model has an open boundary located at approximately 50°E, which is forced with temperature, salinity, sea surface height and velocities conditions obtained from GLORYS2 (version 4), the most recent ocean reanalysis product by Mercator Ocean (http://marine.copernicus.eu/). Data are available for the period 1993 to 2015, with a 1/4° horizontal resolution and 75 vertical layers, forced by the ERA-Interim reanalysis product created by the European Centre for Medium-Range Weather Forecasts.

The model has a horizontal grid spacing of approximately 1 km (1/100°) and 50, non-uniformly spaced vertical layers (z-coordinates), with an exponentially increasing thickness ranging from 4 m near the surface to 250 m in the deepest layer. There are a total of 18 layers in the first 150 m of depth; the increased model resolution was motivated by recent observational (Zhan et al., 2014; Raitsos et al., 2017) and modeling (Chen et al., 2014; Yao et al., 2014a,b) studies that revealed complex circulation patterns occurring at a wide range of scales. The increased resolution also provides an improved representation of the complex bathymetry of the RS, including the shallow and narrow constrictions of the adjacent gulfs, which is important for the reliable representation of the RS circulation (Sofianos and Johns, 2015).

To accurately simulate the circulation (particularly in the two relatively small gulfs in the north) and the small-scale features associated with the complex regional topography (particularly narrow-mountain-gap winds), the RS model was forced by atmospheric fields with a 5 km horizontal resolution. The high-resolution atmospheric dataset was generated by downscaling the ERA-Interim reanalysis (Dee et al., 2011) using the Advanced Research version of Weather Research and Forecasting (WRF) model, and assimilating available in situ and satellite observations of the region. This substantially improved the representation of regional atmospheric features (Viswanadhapalli et al., 2016; Langodan et al., 2017a,b; Hoteit et al., 2021). The atmospheric fields were made available every three hours to account for the high diurnal variability of atmospheric conditions, especially near the coastal regions. Wind stress was derived following the parameterization of Large and Yeager (2004), considering the relative velocity of the atmospheric wind speed to the simulated ocean surface current velocity. The atmospheric flux components were estimated using the Large and Pond (1981) bulk formulas. The vertical distribution of downward shortwave radiation in the water column has been parameterized following Paulson and Simpson (1977), through an exponential decay with depth, assuming water type IB as defined by Jerlov (1968).

The RS MITgcm configuration follows that of Brannigan et al. (2015). A third-order, nonlinear, direct space-time flux limited scheme was used to model the temperature and salinity advection, with scale-selective diffusion and viscosity. Adaptive viscosity schemes were employed with a Smagorinsky coefficient of 3 and corresponding Leith coefficients equal to 1. Biharmonic operators were applied for diffusion and viscosity, for both salinity and temperature. The biharmonic horizontal diffusion coefficient for heat and salt was set to $1 \times 10^7$ m$^4$ s$^{-1}$. In addition, a Laplacian operator with a diffusion coefficient of $4 \times 10^{-6}$ m$^2$ s$^{-1}$ was applied for the vertical mixing of both temperature and salinity. The unresolved upper ocean mixing processes including turbulence induced by wind stress and the associated vertical shear were parameterized using the K-profile parameterization (KPP; Large et al., 1994). The KPP scheme enhances vertical mixing in a boundary layer under the destabilizing influence of surface buoyancy and momentum forcing (Large and Gent, 1999). The KPP scheme further introduces a vertical non-local transport to account for vertical convective mixing. A no-slip boundary condition was applied on the lateral boundaries, while a quadratic drag coefficient of $2 \times 10^{-3}$ was applied at the bottom. A sponge layer of 10 grid boxes with a relaxation period of five days was implemented for the velocity fields at the open boundary. Evaporation in the model was represented as an equivalent virtual salt flux and the velocities at the open boundary were adjusted accordingly to enforce zero net volume fluxes (e.g. Yao et al., 2014a; Hoteit et al., 2021). No tidal forcing was used in the simulation and no tracer relaxation was applied. A time step of 90 s was used to ensure model stability over long-term simulations.

After a five year spin-up run using periodic forcing corresponding to that from the year 2001, the model forced with the downscaled WRF fields was integrated over the period from 2002 to 2015. Daily averaged outputs were stored for all the required diagnostic variables.

2.2. CTD observations

We have analyzed a total of 831 conductivity-temperature-depth (CTD) casts from the RS, spanning a time period of more than 40 years (1982-2013), 551 of which were within the time period of the model simulation (2001-2015). The sources, temporal and spatial coverage of the CTD profiles, and the number of CTD casts, are listed in Table 1. These include observations collected during all the major expeditions that took place in recent years (in addition to those listed in Alraddadi, 2013), which were supplemented herein with observations from several of the most recent expeditions.

**Table 1**. Expedition name, time period, spatial coverage, number of casts and the source of the available CTD observations considered in this work.

|    | **Expedition**      | **Period**  | **Spatial coverage**    | **Num** | **Source**                  |
|----|---------------------|-------------|-------------------------|---------|-----------------------------|
| 1  | R/V Marion Dufresne | Oct 82      | Whole Red Sea           | 77      | Maillard and Soliman, 1986  |
| 2  | R/V Sagar Kanya     | May 83      | Whole Red Sea           | 33      | Quadfasel and Baudner, 1993 |
| 3  | R/V Meteor 5/2      | Feb-Mar 87  | CRS                     | 59      | Quadfasel and Baudner, 1993 |
| 4  | R/V Meteor 5/5      | Jul-Aug 87  | CRS and SRS             | 52      | Quadfasel and Baudner, 1993 |
| 5  | R/V Maurice Ewing   | Feb 99      | NRS and Gulf of Aqaba   | 80      | Plaehn et al., 2002         |
| 6  | R/V Meteor leg 44/2 | Aug 01      | Whole Red Sea           | 77      | Sofianos and Johns, 2007    |
| 7  | KAUST-WHOI          | Oct 08      | CRS                     | 35      | Alraddadi, 2013             |
| 8  | KAUST-WHOI          | Mar 10      | NRS and CRS             | 110     | KAUST repository            |
| 9  | KAUST-WHOI          | Sep 11      | NRS and CRS             | 262     | KAUST repository            |
| 10 | KAUST 2013          | Mar 16      | CRS                     | 46      | KAUST repository            |

The geographical and temporal distributions of these observations are shown in Fig. 2. More observations are available in the eastern part of the RS and along its central axis than in the western part of the basin (Fig. 2 c, d). Considerably more observations were conducted after the period after 2008 (Fig. 2a) and particularly in summer, whereas the number of observations taken in winter is relatively small (Fig. 2b). From 831 (551 during the 2001-2016 simulation period) casts analyzed, 535 (372) observations were collected in the summer period (May–September) and 261 (179) in the winter period (October–April). Availability of data is particularly low in the SRS during winter period (Figs. 2a,d). Despite the sparsity of these observations, they still provide valuable insight into the near-surface stratification, and they have been used to validate the model.

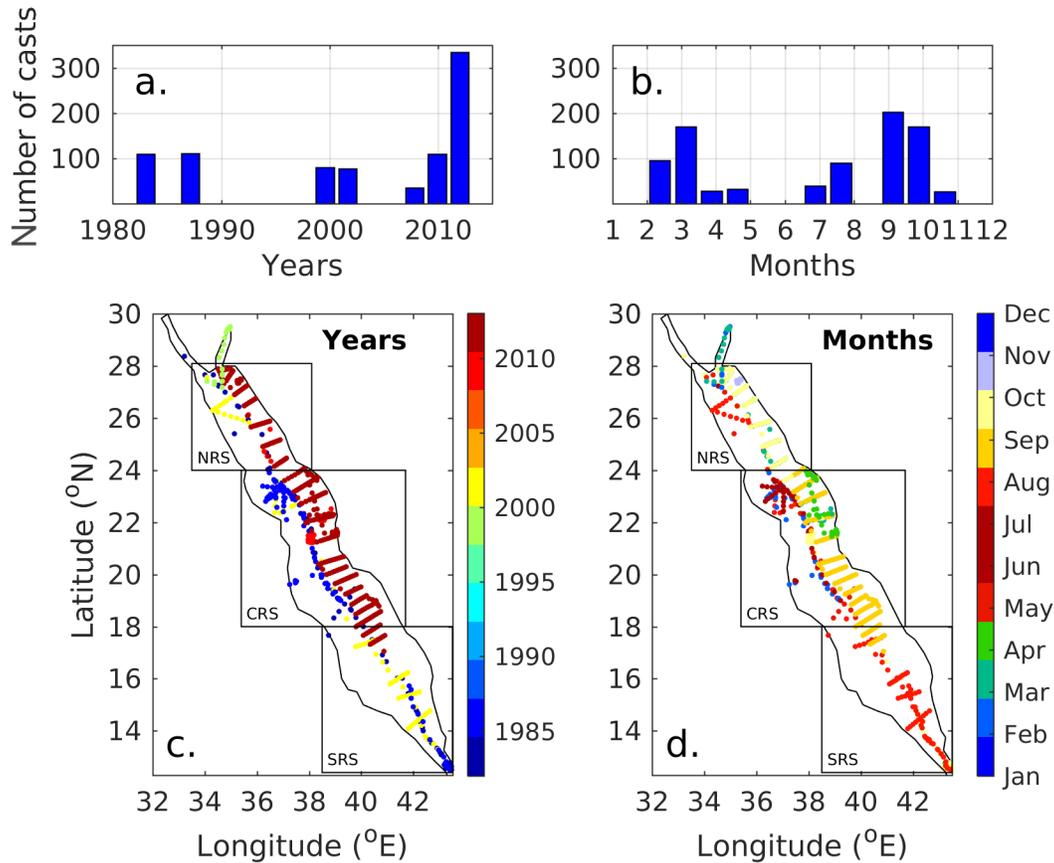

*Figure 2. Number of CTD casts in the RS and their geographical locations per year and per month.*

2.3. Method for determining MLDs

Several methods have been proposed in the literature to provide global estimates of MLD. A common difficulty is that the methods depend on the thermohaline conditions of the regions (e.g., Kara et al., 2003), as well as the type and vertical resolution of the input data (e.g. Huang et al., 2018 and references within). We sought to identify a method that would provide robust MLD estimates using both the high-resolution vertical CTD profiles and numerical model outputs, which has a coarser vertical resolution and is archived as a daily average.

Criteria that have been commonly used to determine the MLD can be separated into two general groups: property difference-based criteria, and gradient-based criteria. In the property difference-based criteria, the MLD is defined as the depth at which the vertical change in some oceanic property (most commonly the potential temperature or potential density) from the surface value at a particular horizontal location exceeds a threshold value (e.g., Montégut et al., 2004). Lorbacher (2006) suggested a gradient-based criterion whereby MLD is determined as the shallowest extreme curvature of the near surface layer density or temperature profiles. However, since the curvature method relies on considering the second derivative, it may fail to provide accurate MLD estimates when dealing with "wiggly" profiles (Lorbacher et al., 2006).

Huang et al., (2018) suggested a new method that provides a universal means of identifying the MLD. This method is based on the idea that density (as well as temperature and salinity) varies more strongly as we approach the MLD. Accordingly, the authors defined a relative variance as the ratio between the standard deviation of the variable's profile and the maximum variation over that depth range. At the depth of the ML the maximum variation increases sharply and is stronger than the standard deviation over this depth range. An initial estimation of a point residing below the ML depth is therefore possible by identifying the minimum of the relative variance. A final adjustment is then made by examining the rate of change of standard deviation above this depth. Huang et al. (2018) provided an extensive comparison with all commonly used methods from both families and further demonstrated its robustness for identifying the MLD in noisy data. In the present study, we used the density profiles to identify the MLD, because in the RS neither temperature nor salinity can globally define the ML. The vertical profiles of properties in the RS vary regionally owing to differences in the prevailing atmospheric forcing, as well as the strong advective fluxes of water exchanges at the straits connecting the RS to its adjacent Gulfs.

For a thorough evaluation of the accuracy of the different methods, we evaluated four criteria for determining the MLD in the RS. Two of these were based on property differences (density and temperature thresholds), one based on the gradient of the density profiles and the one proposed by Huang et al. (2018). For the difference-based criteria a threshold value of $\Delta\sigma=0.125$ kg/m$^3$, which is in the range of commonly used thresholds (e.g., Dong, 2008), was chosen as the most appropriate for the RS. Accordingly, for the temperature-based method, the threshold was chosen as the potential temperature difference equivalent to a potential density increase of $\Delta\sigma=0.125$ kg/m$^3$.

The available observations were first examined for their proper representation of the local ML. Estimation of MLD requires gap-free profiles that are also sufficiently deep to capture the vertical ML structure. From 554 profiles available from the model simulation time period, 51 profiles did not satisfy these criteria. The final comparison was performed using 322 profiles from the summer period, spanning all the regions of the main Red Sea basin (excluding the northern gulfs) and 181 profiles from the winter period, mainly from the NRS and CRS regions. No observations were available from the GoAq and GoS during the model simulation time period in either of the two seasons.

We next assessed the robustness of the four methods using both the model results and the observations. The corresponding results from the simulation and the observations were individually examined and the estimates were compared with visually identified MLDs. Results for several selected profiles are shown in Fig. 3. The most robust MLD estimates were obtained using the relative variance method proposed by Huang et al. (2018). The method was able to correctly identify the MLD in the majority of the profiles (in 93% of the profiles, compared to 85% for the curvature method and less than 80% for the threshold methods) and performed well for different vertical structure's characteristics. It was able to account for the different characteristics of the water column along the RS (i.e., whether it was temperature or salinity dominated), as it does not depend on subjective thresholds, which are usually tailored to specific conditions. This is illustrated in Fig. 3a, b. Although this method resembles the gradient approach in that it searches for the depth of maximum variability, it has generally been more robust, showing less sensitivity to noisy data or weakly stratified profiles. This was especially important for the in situ datasets, as illustrated in Fig. 3c, d. The method proposed by Huang et al. (2018) was therefore used to estimate the MLD from both observations and model outputs, as described in the next section.

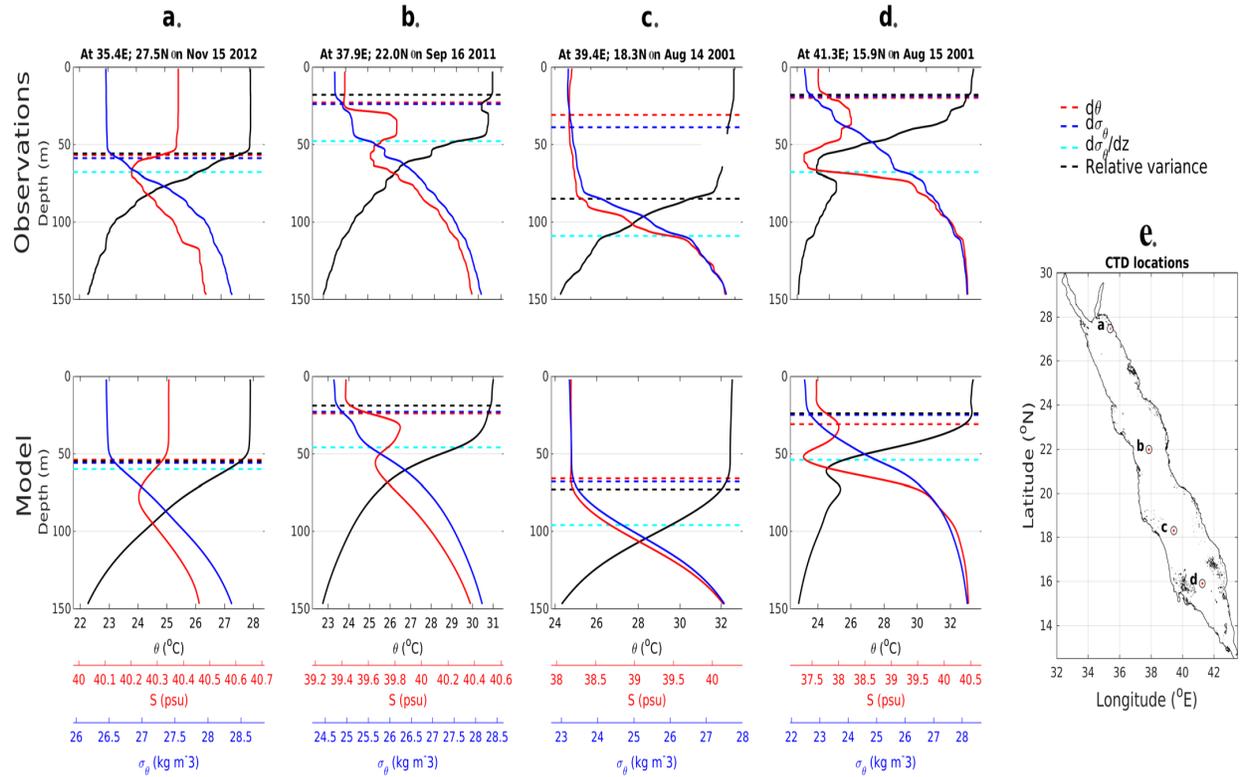

*Figure 3. MLD estimates for selected profiles of salinity S, potential temperature θ, and potential density $\sigma_\theta$, obtained using the following four methods: two property difference-based methods, with $\sigma_\theta$ threshold of $0.125\ kg/m^3$ and $\theta$ threshold that is equivalent to the $\sigma_\theta$ difference of $0.125\ kg/m^3$; a maximum gradient method based on the $\sigma_\theta$, and the relative variance method of Huang et al. (2018), also applied on $\sigma_\theta$.*

## 3. Validation of the simulated near surface stratification

The near-surface (upper 350 m) model potential temperature ($\theta$), salinity (S) and potential density ($\sigma_\theta$) have been validated against the CTD profiles. Additional model validation is provided in the Supplementary Material, including the model's representation of the sea surface temperature (Section S2) the sea-level variability (Section S3), and the circulation patterns (Section S4).

### 3.1. Comparison with CTD observations

The mean values and standard deviations of the vertical profiles of the potential temperature $\theta$, salinity S and potential density $\sigma_\theta$ in the top 350 m, as obtained from the model and the CTD observations, are presented alongside their respective root-mean-square differences (RMSDs) in Figs. 3 and 4. The profiles and corresponding RMSD values were temporally averaged over the individual RS regions and over October to April, most representative of the winter and May to September, representive of the summer periods. Following Raitsos et al. (2013) and Ahmad and Albarakati (2015), we separately consider the two gulfs in the north (GoAq and GoS), the NRS (24º–28ºN), the CRS (18º–24ºN), and the SRS (between 18°N and the BAM), as shown in Fig. 1. This partitioning is based on the mean $\theta$ and S characteristics and the unique bathymetric features of each region. The model results were validated using the CTD observations within the time period of the model simulation (2001-2015); however, no observations were available in GoAq and GoS, and in the SRS in winter during the simulation period. In these cases we compared the available observations with the modeled climatological means for the same location and season. The mean vertical profiles from all the available CTD observations are shown in the Supplementary Material (Fig. S6).

*Gulf of Aqaba.* In the GoAq 61 sample stations were available from one expedition in winter (Feb 21– March 5, 1999). They showed the lowest surface temperatures and the largest MLD (more than 300 m) in the RS. Comparison with the model climatological means (Fig. 4a), suggests that both are representative of typical winter conditions (Carlson et al., 2014).

*Gulf of Suez.* Only two sample stations were available in October 1982, showing the typical strong salinity stratification that develops at the end of the summer period, while temperature shows a smaller vertical gradient. Salinities at depths exceeding 30 m are the highest observed in the RS, whereas the water above is less saline (Fig. 4b), due to the surface inflow from the NRS (Sofianos and Johns, 2017). This vertical salinity profile agrees well with the modeled climatology, whereas the observed temperatures lie within the range of the simulated climatological mean temperature profiles.

*North Red Sea.* In the NRS, 84 observations were available during the summer and 89 during the winter period, providing relatively good coverage (Fig. 5a, b). The simulated temperature and salinity closely match the vertical profiles based on the observations, as well as their seasonal variability (Fig. 5a, b). The RMSDs for all properties are small (<0.1 °C for $\theta$, < 0.2 psu for S and < 0.25 for $\sigma_\theta$ in both seasons), and remain relatively constant with depth. A small increase in the RMSD occurs in summer at the depth of the thermocline, where the vertical gradients of both temperature and salinity are strongest (Fig. 5b). Nevertheless, the depth of the homogenized layer in the model and the observations are in good agreement.

*Central Red Sea.* The largest number of observations is available from the CRS (216 in summer and 90 in winter, Fig. 5c,d). During the winter period the modeled and observed mean properties are very similar, with RMSDs less than 0.2°C for $\theta$, 0.3 psu for S and 0.3 for $\sigma_\theta$, everywhere within the top 350 m (Fig. 5c). The agreement is particularly good in the top 100 m. In summer, the thermocline shoaling to a depth of less than 30 m is accurately reproduced by the model. Differences are larger within the intermediate depth range (50–150 m), where the observations show a local temperature minimum, which is even more pronounced in salinity (Fig. 5d). These are typical summer conditions caused by the advection of colder and fresher GAIW from the SRS (Sofianos and Johns, 2007). The model reproduces the observed increase in variability of both temperature and salinity over the same depth range (50–150 m), although the modeled variability is somewhat weaker, resulting in higher RMSDs in this depth range. This may be due to differences in the properties of the GAIW inflow and/or too strong mixing of the GAIW in the model.

*South Red Sea.* In the SRS, 72 CTD observations are available in summer, while only three observations are available in winter and they fall outside of the model simulation period. Although the number of observations is small, they show a large seasonal variability that is characteristic for this region (Fig. 5e,f), caused by the seasonal reversal of the monsoon-driven water exchange with the GoAd (Sofianos and Johns, 2015).

In winter, the water in the approximately top 130 m is warmer and fresher owing to the inflow of surface waters from the GoAd, which overlie colder and saltier RS intermediate waters (Fig. 5e). The model's climatological means agree well with the main characteristics of these two water masses and the depths of their transition.

In summer, the model reproduces the strong vertical stratification with a thin, warm and highly saline vertically homogenized upper layer (approximately 25 m deep) overlying a colder and fresher layer that extends to approximately 300 m depth (Fig. 5f), caused by the summer intrusion of the GAIW (Churchill et al., 2014; Dreano et al., 2016). The model reproduces the depth of the transition to the intermediate GAIW layer, showing the strongest variability of properties in the 50-250 m depth range; however, the minimums of salinity and temperature are slightly underestimated, as was the case in the CRS. This yields larger RMSDs for both temperature and salinity in the depth range of the intermediate waters. However, the differences in the temperature and salinity have opposite effects on the density, leading to a relatively better representation of the vertical density stratification.

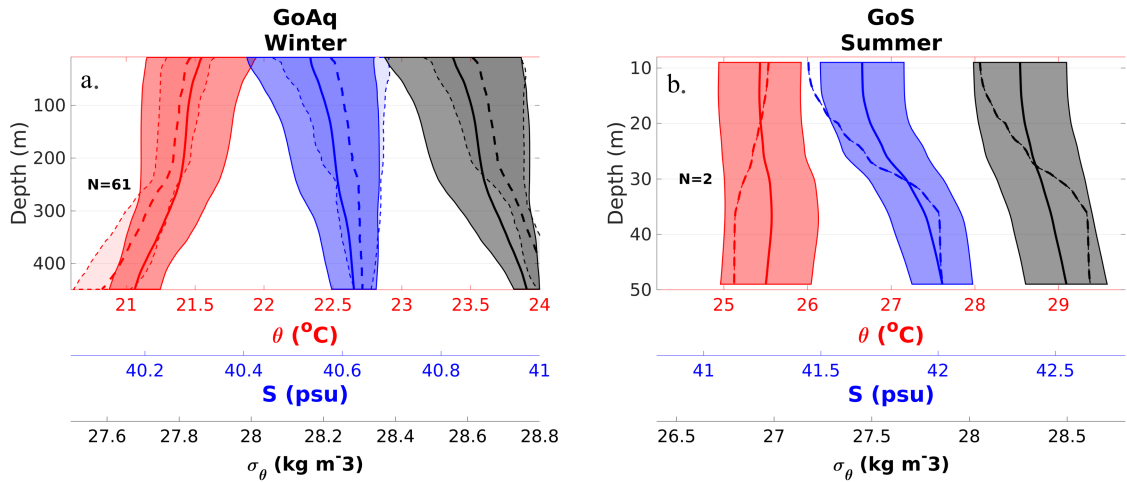

*Figure 4. Mean vertical profiles of salinity S, potential temperature θ, and potential density $\sigma_\theta$, obtained from CTD observations (dashed lines) and model results (solid lines) over the GoS in summer and GoAq in winter; note different scales. Shading shows the corresponding standard deviation and N denotes the number of observations considered.*

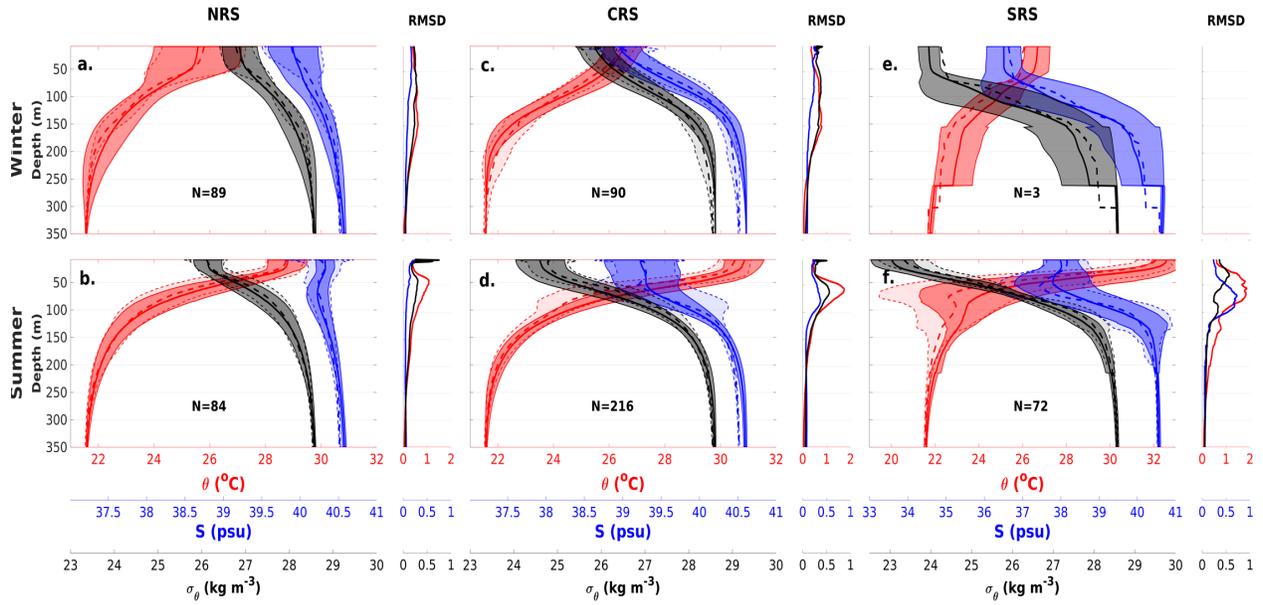

*Figure 5. Mean upper layer (top 350m) vertical profiles of salinity S, potential temperature θ, and potential density $\sigma_\theta$, from CTD observations available during 2001-2015 (dashed lines) and the model results (solid lines), and the corresponding RMS differences, over the (a,b) NRS, (c,d) CRS and (e,f) SRS regions (as in Fig. 1). (a,c,e) and (b,d,f) correspond to winter and summer respectively. Shading shows the corresponding standard deviation and N denotes the number of observations considered.*

### 3.2. Comparison of MLD estimates from the model and CTD observations

The modelled MLD estimates were validated against those derived from the CTD observations for the same location and same day. They show generally good agreement during both seasons and throughout the basin (Fig. 6) despite the differences in the vertical and temporal resolution between the model outputs (daily averages; vertical resolution ranging from 4 m at the surface to 14 m at the maximum MLD estimate) and in situ observations (instantaneous; usually with a vertical resolution of 1 m).

During the summer period (Figs. 6a, b), MLD estimates based on both the measurements and modelling mostly range between 20 and 40 m, with very few estimates deeper than 60 m. During the winter period (Figs. 6c, d), the model reproduces the observed ML deepening (40–60 m), except NRS and GoAq, where no measured data were available for comparison.

During the summer period, MLD estimates show smaller differences along the basin's central axis (Fig. 6e). The largest differences are in the northwestern part of the RS during winter (Figs. 6f), where MLD estimates from both observations and model results are the deepest (not shown). The RMSD is 13.8 m (12.6 m during the summer period and 15.6 m during the winter period), with differences showing a uniform spread in both seasons (Fig. 6g). In the CRS, the differences do not exhibit a clear spatial pattern. The MLD estimates obtained from the few observations available for the NRS in winter (where most of the deep mixed layers are observed, Fig. 6f) agree well with the model estimates.

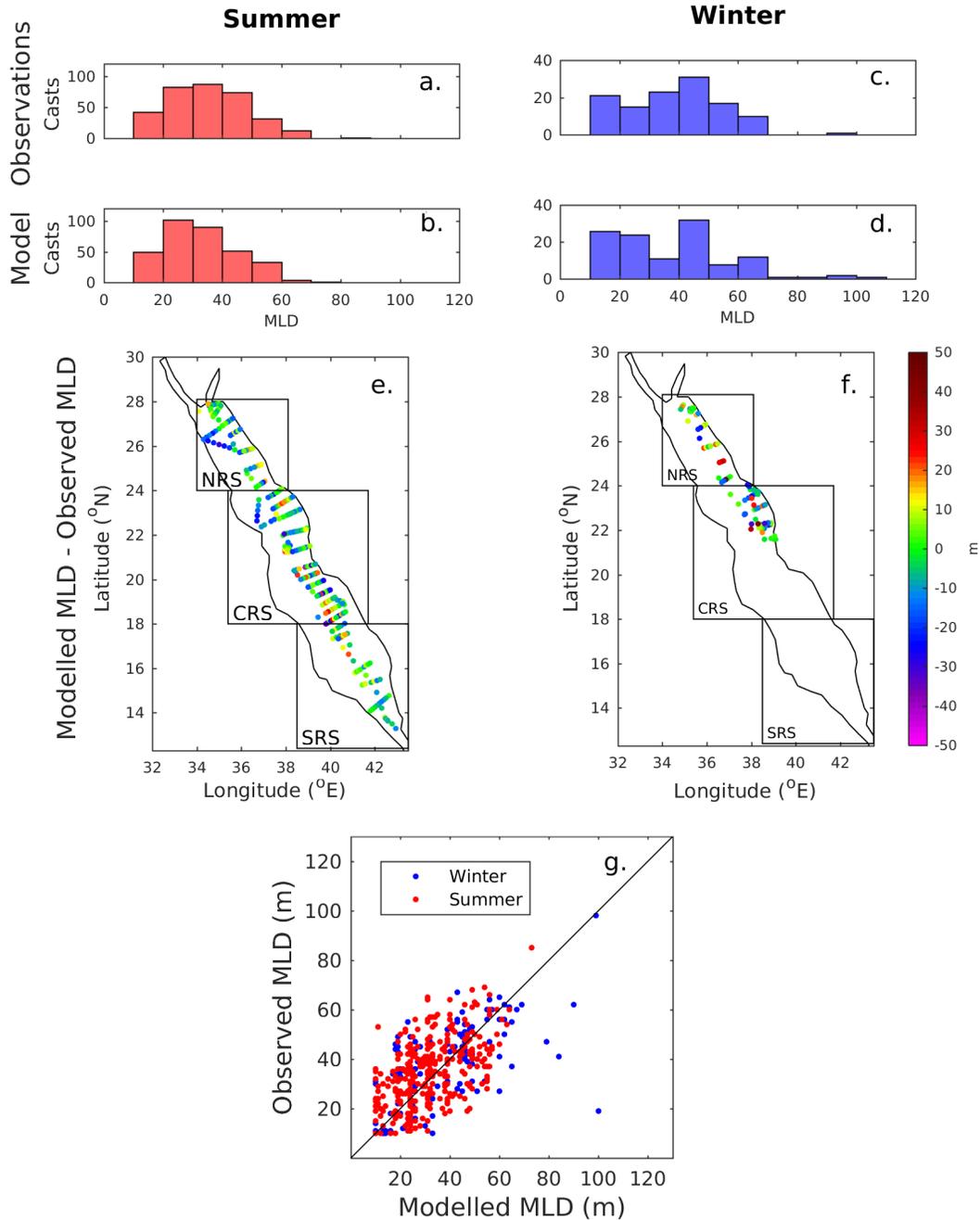

*Figure 6. (a, c) Histogram of the number of CTD casts available during the model simulation period (2001-2015) and (b, d) from daily averaged model output, shown as a function of MLD; (e, f) the difference between the MLD estimates obtained from the CTD casts and the daily averaged model output; (g) scatter diagram of the MLD estimates.*

## 4. Atmospheric buoyancy and momentum fluxes

As surface momentum and buoyancy fluxes are generally the major drivers of mixing in the near-surface layer and the oceanic ML development (Cronin & Sprintall, 2008), we hereby briefly describe the mean seasonal characteristics of wind stress and surface buoyancy flux. We also separately consider the heat and freshwater flux components, as they generally exhibit different spatial and seasonal variabilities. The relative importance of the surface buoyancy and momentum fluxes on oceanic turbulent mixing that ultimately leads to the development of the ML, is in addition qualitatively assessed by examining the Obukhov length scale.

**4.1.** Wind forcing

The seasonal climatology of wind stress in the RS for summer (June–August), autumn (September–November), winter (December–February), and spring (March–May) is presented in Fig. 7. The wind fields used to force the model simulation in this work have been analyzed in Langodan et al. (2017). As the spatiotemporal characteristics of wind stress closely resemble those of wind fields, we here briefly discuss only the most relevant features.

In winter, wind stress is intensified due to strong northerly winds in the NRS and CRS, that are stronger in the western parts of the basin (Papadopoulos et al., 2013). Channeling of winds by topography intensifies wind stress along the axes of the GoAq and GoS in the southwesterly direction, strengthening them toward the southern parts of the two gulfs. In the NRS and CRS wind stress is stronger in the western parts of the basin. Strong southeasterly winds induce wind stress in the SRS especially near the BAM. They blow northward with progressively lower intensity, mostly influencing the western parts of the SRS and CRS (Pratt et al., 1999; Langodan et al., 2017).

In summer, wind stress is intensified by locally strong winds that develop in the vicinity of gaps in the coastal mountain chains parallel to the RS axis (Jiang et al., 2009; Menezes et al., 2018). Of particular importance are eastward wind jets that blow through the Tokar Gap in the CRS, before shifting southward towards the eastern part of the SRS. Winds stress in the SRS is generally stronger in the vicinity of the BAM due to channeling of winds through the strait.

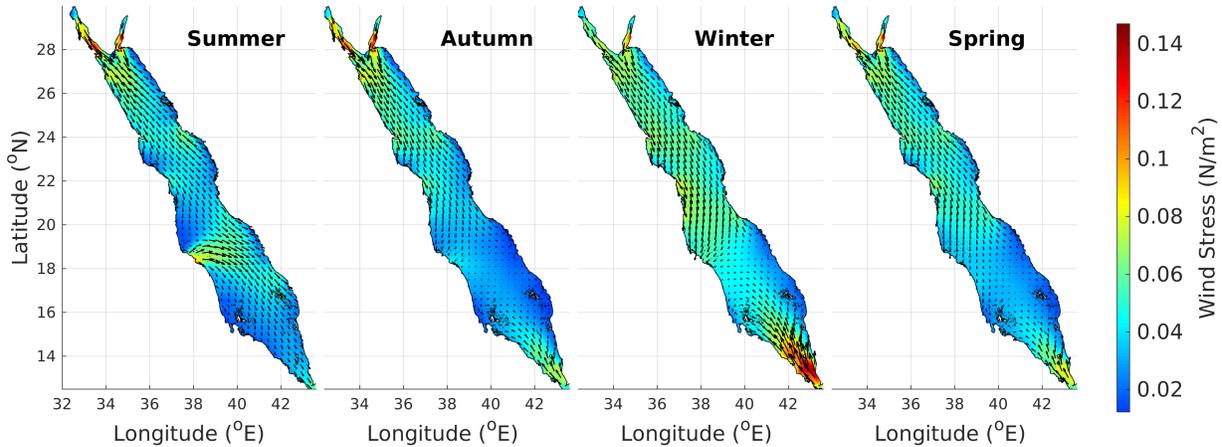

*Figure 7. Seasonal climatology of the model estimates of wind stress.*

4.2. Surface buoyancy fluxes

The net surface buoyancy flux ($B$) is the sum of the surface heat and freshwater (FW) flux components and it can be expressed as

$$B = -g \cdot a_t \cdot \frac{Q}{\rho_s \cdot c_p} + g \cdot b_s (E - P) \cdot S, \qquad (1)$$

where $\rho_s$ is the water density at the surface, g= 9.81 ms$^{-2}$ is the gravitational constant, $c_p$ is the specific heat capacity, $a_t$ and $b_s$ are the thermal expansion and saline contraction coefficients[*] evaluated at the sea surface, and $S$ is the surface salinity; Q is the net heat flux, which is the sum of the net shortwave and longwave radiation, and the sensible and latent heat fluxes. Finally, $E$ denotes evaporation and $P$ precipitation. All terms were evaluated using the daily mean model outputs. Positive buoyancy fluxes increase the stability of the water column and inhibit vertical mixing, whereas negative buoyancy fluxes reduce stratification and drive convective mixing.

    The seasonal climatology of the net buoyancy flux and its heat and freshwater flux components are presented in Fig. 8. The heat and FW fluxes exibit different seasonal variabilities and spatial patterns (Fig. 8). The seasonal variability of the net buoyancy flux (Fig. 8a) is mainly controlled by its heat flux component (Fig. 8b), while the seasonal variability of FW flux component is much smaller (Fig. 8c). Evaporation persists throughout the year, reducing the near surface stratification and playing an important role in preconditioning for ML deepening (Sofianos and Johns, 2015).

    Surface buoyancy loss in the RS starts in early autumn, driven predominantly by surface cooling (Fig. 8a). Both heat and FW fluxes exhibit a north–south gradient, with stronger heat and FW loss in the NRS (Fig. 8b, c). The strongest wintertime surface buoyancy loss occurs in the GoAq, where the monthly mean heat loss can exceed 300 W/m$^2$, and evaporation can be as strong as 3 my$^{-1}$. Although the GoS is located at similar latitudes as the GoAq, the wintertime buoyancy loss is significantly weaker, so that the monthly mean heat losses remain lower than -150 W/m$^2$ and FW fluxes less than -1.5 my$^{-1}$. These differences are caused by the smaller thermal capacity of the GoS, owing to its shallower depth. Wintertime buoyancy loss is also strong in the northwestern part of the NRS, where the monthly mean heat fluxes during winter exceed -250 W/m$^2$ and FW fluxes -2,5 my$^{-1}$. They are driven by cold air outbreaks that originate from the northwest and by the channeling of northern winds from the two gulfs. Locally, increased buoyancy loss near the coasts in the CRS and NRS is mostly driven by the strong surface FW loss (Fig. 8a,c), and mainly through evaporation due to lateral dry winds that blow through mountain gaps. In contrast to the rest of the RS, the sea surface in the central part of the SRS gains buoyancy even during winter (Fig. 8a), when warmer southerly winds isolate it from the cooler atmospheric systems of the north (Langodan et al., 2017).

    In spring, most of the Red Sea gains buoyancy by heating from the atmosphere, especially in the SRS. Buoyancy loss is simulated only in the northern RS, in the vicinity of the straits connecting to the two gulfs, which persists also during the summer period. (Fig. 8a). In summer, buoyancy loses intensify in the region influenced by the Tokar Jet, due to extreme evaporation (Fig. 8a,c). The strongest heat (and buoyancy) gain in summer is simulated in the eastern part of the NRS and in the southeastern SRS (Fig. 8a, b).

---

[*] Coefficients $c_p$, $a_t$ and $b_s$ were computed using the Gibbs-Sea Water (GSW) Oceanographic Toolbox (McDougall and Barker, 2011)

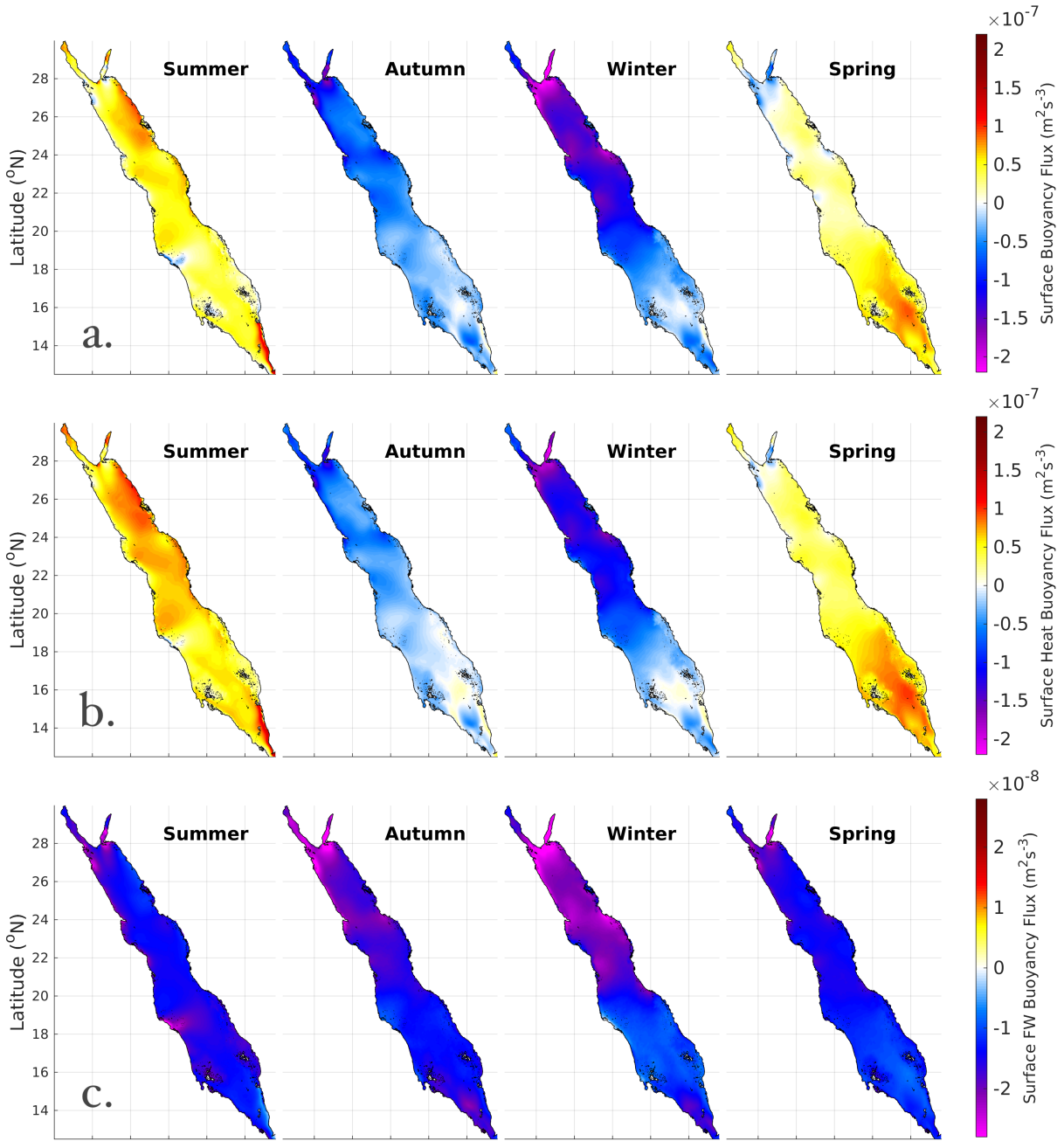

*Figure 8. Seasonal climatology of the model estimates of (a) net air–sea buoyancy flux, and its (b) heat flux and (c) freshwater flux components. Positive values indicate ocean surface buoyancy gain.*

4.3. The Relative Importance of surface buoyancy and momentum forcing on vertical mixing

The development of the ML is primarily driven by vertical mixing, which is generated by both wind-induced turbulence, through the shearing effects of momentum forcing, and by convective overturning, driven by the sea surface buoyancy loss. To assess the relative contribution of wind stress and buoyancy forcing to vertical mixing, we use the Obukhov length scale (Obukhov, 1946, 1971; Foken 2006). The Obukhov length $L_m$ is a universal length scale for exchange processes in surface layers and represents the ratio of the buoyancy and shearing effects (Phillips, 1977; Large, 1988). Therefore, it is a measure of the relative importance of wind and buoyancy forcing on oceanic turbulent mixing that lead to the development of the ML (Taylor and Ferrari, 2010). The length scale $L_m$ is defined by

$$L_m = \frac{(U^*)^3}{\kappa B_o},$$

where $U^*$ is the friction (or shear) velocity, $\kappa = 0.4$ is the von Karman constant and $B$ is the buoyancy flux. The friction velocity represents the contribution of shear to turbulence. The buoyancy term is a measure of the contribution of convective mixing to turbulence, when buoyancy flux is negative, or turbulence suppression, when buoyancy flux is positive (Large, 1998). The estimates of the length scale $L_m$ represent the depth at which turbulence is dominated by buoyancy than by wind induced shear (Taylor and Ferrari, 2010). The surface buoyancy flux is estimated as in Eq. (1) (Section 4.2), considering the contributions of its components inside the ML, while the friction velocity $U^*$ is defined by

$$U^* = \sqrt{Ws/\rho_s}, \qquad (B2)$$

where $Ws$ is the wind stress and $\rho_s$ is the surface density.

Because of the nonlinear nature of $L_m$, local extremes may appear in the regions in which $B$ changes sign and its value approaches zero. As these values indicate the absence of buoyancy forcing, they were removed using a maximum threshold for $L_m$ (<150 m) based on the examination of the distribution of $B$ in daily model outputs, thereby ensuring that other regions were not affected.

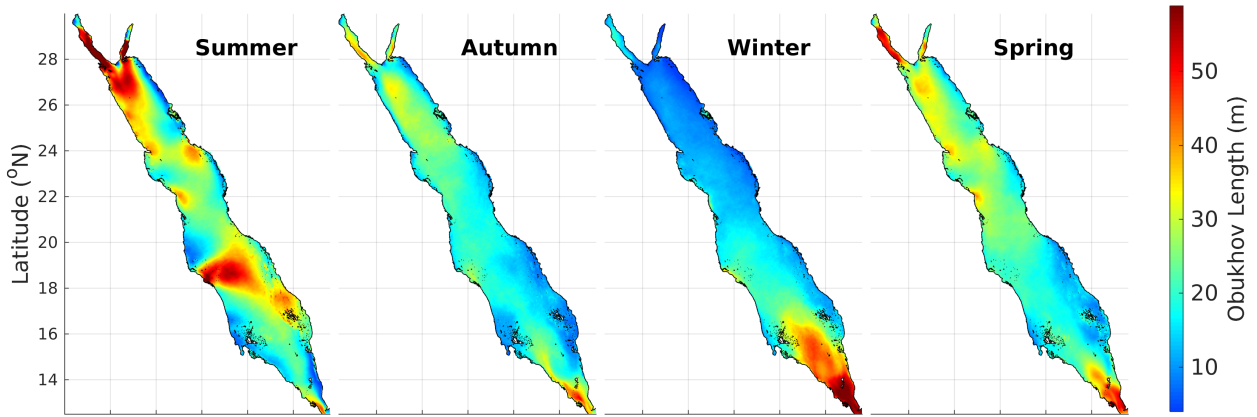

*Figure 9. Seasonal climatology of the Obukhov length scale.*

The dominant feature of the seasonal climatological means of $L_m$ obtained from the 2001-2015 model time average is strong spatiotemporal variability. Nevertheless, characteristic seasonal and spatial patterns emerge (Fig. 9). The seasonal variability is mostly controlled by the variability of the surface buoyancy fluxes. The strongest buoyant mixing (small $L_m$) is in the northern Red Sea and GoAq in winter, associated with convection and dense water formation (Fig. 9). Wind induced mixing has a higher contribution in the other seasons, especially in summer, when buoyancy acts to increase the stratification. Its influence decreases in winter when convective processes dominate vertical mixing.

Despite the dominant role of buoyancy forcing in the seasonal variability, the spatial patterns of $L_m$ generally match the wind stress variability. The strongest wind induced mixing is driven by channeling of winds through the straits (at the two Gulfs in the North and at the BAM), and strong wind stress from jets blowing through mountain gaps. In summer, wind forcing dominates over convective forcing in the northern parts of the basin, as northern winds are channeled through the narrow Gulfs and exit towards the NRS. Wind stress is also especially strong in the western CRS in the Tokar jet region, and its effects extend to the western part of the SRS, especially over the shallow regions. During autumn, the wind contribution gradually decreases, as buoyancy loss starts to increase everywhere in the RS, especially in the shallow regions of the SRS. In winter, the wind driven mixing remains strong only in the SRS due to the monsoon-driven southern winds that are channeled through the BAM strait.

## 5. Seasonal evolution of the MLD

The modeled MLs in the RS reveals a strong seasonal and spatial variability that generally resemble the variability of surface forcing, but also reveal spatial patterns related to the circulation features. Maps of the seasonal climatologies of the simulated mean and maximum MLDs are presented in Fig. 10, while monthly spatial averages over the individual regions are presented in Fig. 11. Monthly climatological maps of the simulated MLDs are provided in the Supplementary Material (Fig. S7). The influence of atmospheric buoyancy flux and wind stress on the seasonal MLD evolution is examined by analyzing their daily correlations (Figs. 11a and 11b, respectively).

In summer, the MLs are shallow almost everywhere in the RS (Fig. 10), with the exception of a local deepening in the southern part of the CRS in the vicinity of the Tokar Jet (Fig. 10). MLs start to deepen in mid-September throughout the basin (Fig. 11). Throughout the RS, the deepest MLs occur in January and February (Fig. 11). The MLs are deepest in the northern parts of the basin, particularly in the GoAq and the western parts of the NRS, and generally shoal toward the south, with deeper MLs in the eastern parts of the CRS and SRS. Monthly means over the NRS reach 65 m, but remain at about 20 m in the SRS (Fig. 10). Maximum depths reach locally reach more than 300 m in the NRS and 80 m in the SRS. The deepest MLDs occur in the GoAq, with area monthly means of more than 120 m (Fig. 10a), locally exceeding 400 m depth (Fig. 10b). In early spring, the near-surface starts to restratify and consequently MLs start to shoal throughout the basin, with mean depths generally below 30 m (Fig. 11). Relatively deeper MLs, are sustained in the northern parts of the basin (locally reaching more than 100 m depth) and in the SRS near the BAM (locally reaching 60 m depth) (Fig. 12a).

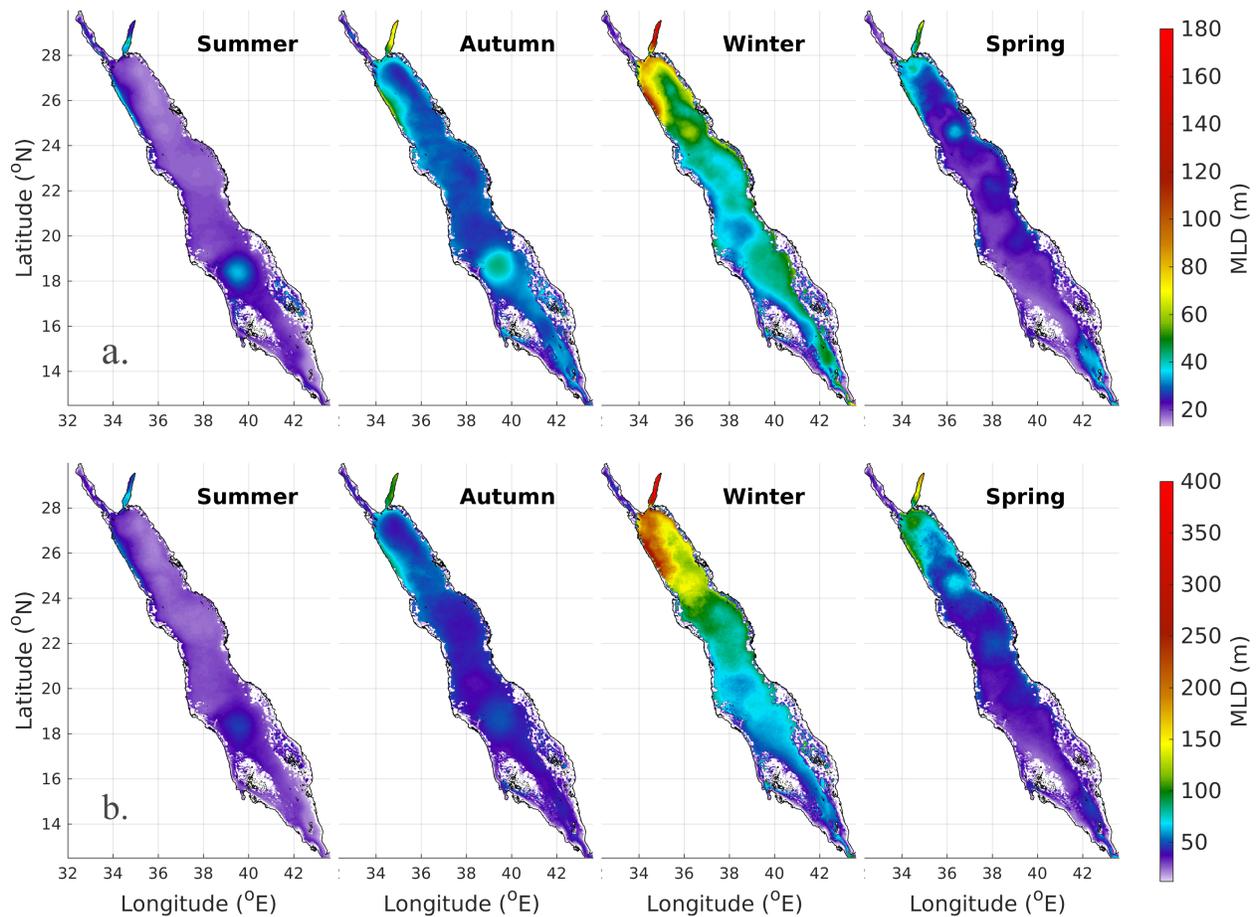

*Figure 10. Seasonal climatology of the (a) mean and (b) maximum model MLD estimates.*

The daily correlations of MLD and surface buoyancy and momentum forcing suggest that the large-scale seasonal variability of MLD is mostly governed by that of surface buoyancy flux. In winter the correlations with the buoyancy forcing are uniformly high (>0.7) in the NRS and CRS (Fig. 12a), where MLs are deepest (Fig. 10). In spring, when the thermal stratification causes the MLs to shoal almost everywhere in the RS, this correlation decreases to approximately 0.3. In summer, the correlations with the wind stress are relatively high (>0.8) everywhere in the RS (Fig. 12b), while the correlations with buoyancy forcing is strong (>0.8) in regions where lateral winds drive strong latent heat loss. In autumn, with the onset of the surface buoyancy loss in the northernmost part of the RS, the correlations with the buoyancy forcing increase in the two gulfs in the north and in the north and western parts of the NRS, while correlation with winds is higher in the western parts of the RS. Unlike in the rest of the basin, in the SRS, and especially in the vicinity of the BAM, the correlation with the wind stress remains high (>0.8) throughout the year.

The MLD in the RS generally shows high correlation with the atmospheric forcing, in both its temporal and spatial variability. However, periods when correlations of both components of atmospheric forcing with MLD are low suggest that other processes also influence its development. For example, the stratification of the water column can also be influenced by lateral transport of heat and salt associated with the general and mesoscale circulation. The northward upper layer flow associated with the general thermohaline circulation in the RS transports warmer waters from the south that tend to increase stratification. Moreover, the presence of mesoscale eddies may affect the MLD locally by increasing mixing or by vertically displacing the isopycnals. We next discuss the main regional MLD characteristics in detail.

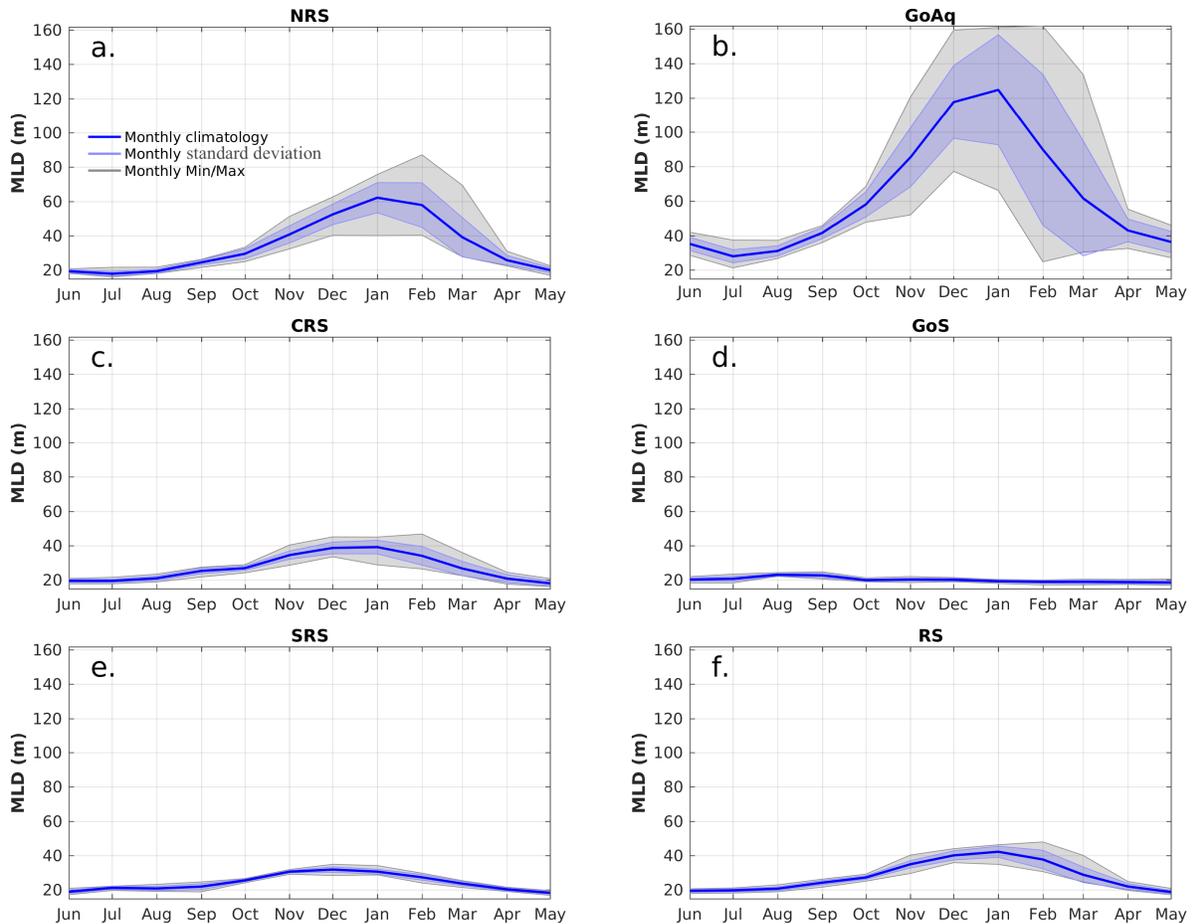

*Figure 11. Monthly mean climatology of the simulated MLD during 2001–2015, along with their monthly standard deviation and the range between their minimum and maximum monthly mean values, averaged over the individual regions: (a) NRS, (b) GoAq, (c) CRS, (d) GoS, (e) SRS, and (f) the entire RS (as in Fig. 1).*

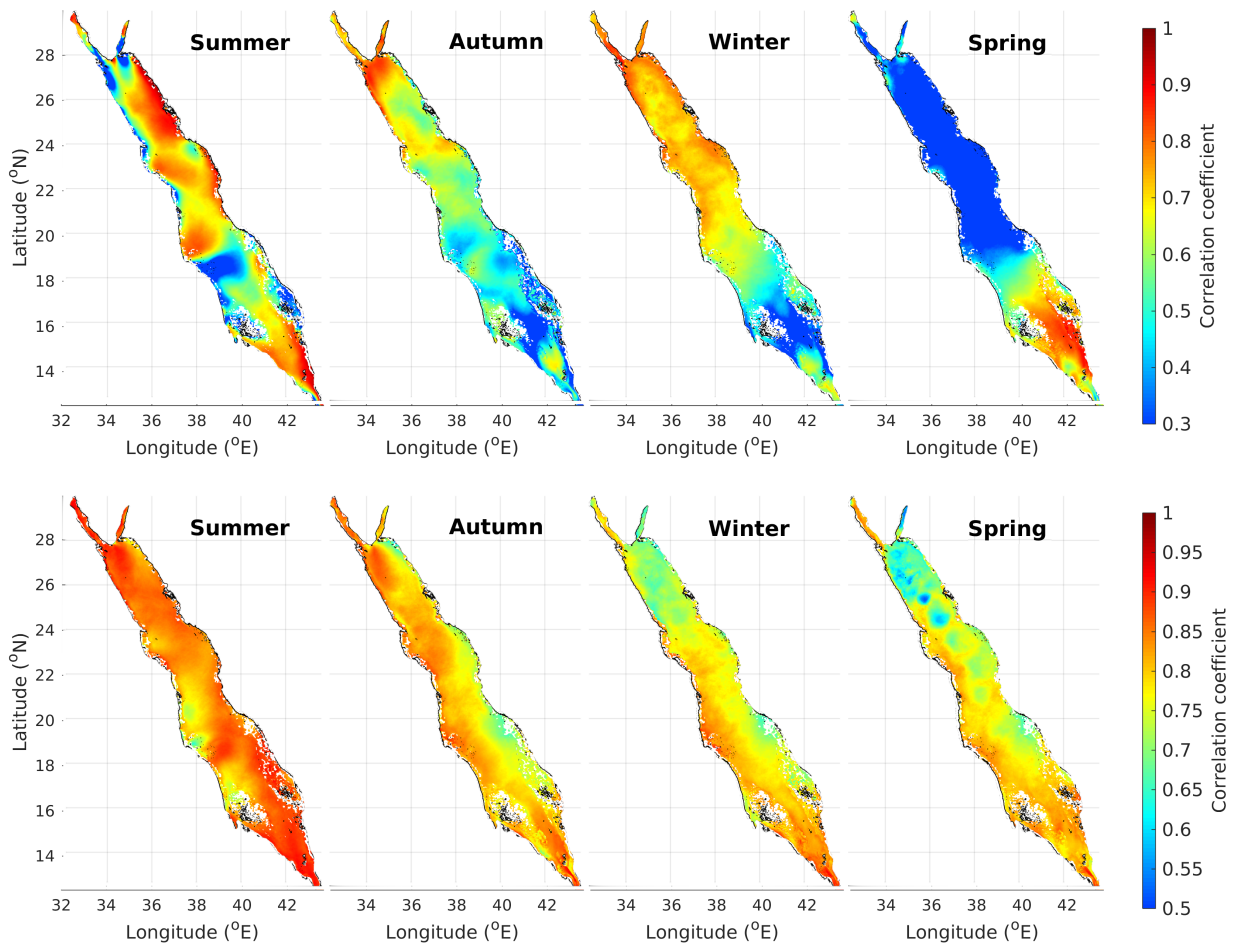

*Figure 12. Daily correlation between the model estimates of MLD and (a) the net air–sea buoyancy flux and (b) the wind stress, respectively.*

*Gulfs of Aqaba and Suez.* Throughout the year, the deepest MLs in the RS develop in the GoAq (Fig. 10b). In winter, the monthly mean MLD averaged over this region can exceed 160 m (Fig. 11), and these deep MLs persist longer than those in the rest of the basin (Fig. 11b). The monthly mean MLDs in the GoAq exhibit large standard deviations indicating that significantly deeper MLs develop in some years (Fig. 11b). MLD during winter shows the highest correlation with surface heat loss (Fig. 12a), indicating that it is the dominant process driving the ML deepening. The deepest MLs are found in the northern part of GoAq, with maximum MLDs locally exceeding 400 m (Fig. 10b), although buoyancy forcing and wind stress are more intense in its central and southern parts (Fig. 7 and Fig. 8, respectively). This is likely a result of the upper layer inflow of warmer and less saline waters from the NRS. They increase the stratification in the southern part of the gulf, inhibiting the development of deep MLs, but their influence reduces towards the north, as has also been suggested in previous studies (Wolf-Vecht et al., 1992; Plähn et al., 2002). In summer, the monthly mean MLD averaged over the GoAq is approximately 30 m deep. Deeper MLs (greater than 50 m) can develop in the southern parts of the gulf (Fig. 10b), where the MLD shows high correlation with strong winds channeled through the Straits of Tiran, suggesting the dominance of wind-induced mixing (Fig. 12b).

The GoS has relatively shallow MLDs and small seasonal variability, because its shallow depths preclude ML deepening. Mixing is mainly driven by winds, except in winter, when buoyancy forcing is relatively stronger, also showing higher correlation with the MLD (Fig. 12a,b). However, MLDs appear slightly shallower in winter than in summer, suggesting that atmospheric forcing is balanced by surface inflow of relatively warmer and fresher waters from the NRS following the intensification of baroclinic circulation during winter (Sofianos & Johns, 2017).

*North Red Sea.* MLDs in the NRS show similar seasonal variability to the GoAq, with deepest MLs in January and February (Fig. 11a), driven mainly by strong wintertime air–sea buoyancy loss (Fig. 12a). ML deepening starts in autumn, and correlation of the buoyancy flux with the MLD generally increases (Fig. 12a), following the onset of buoyancy loss (Fig 8a). The monthly MLD averaged over the NRS can exceed 150 m in winter, also exhibiting relatively large standard deviations that indicate significant interannual variability (Fig. 11a). Deeper MLs develop on the periphery of the NRS (Fig. 10a), reaching maximum depths of more than 250 m (Fig. 10b). The NRS is characterized by a general cyclonic circulation that prevails throughout the year and intensifies in winter (Zhan et al., 2014). In the center of this cyclonic gyre, the MLs are shallower owing to the doming of the isopycnals. MLDs are also deeper in the western parts of the NRS throughout the year, especially during winter months (Fig. 10a). Several factors may contribute to this variability. First, wind forcing is consistently stronger in the western parts throughout the year (Fig. 7). In autumn and winter, the deepest MLs are correlated with the regions of greatest buoyancy fluxes (Fig. 12), which also occur in the western parts of the basin (Fig. 8a). Finally, the eastern parts of the NRS is directly influenced by relatively warmer and fresher waters that are advected from the south in the upper layers following the general cyclonic circulation. These waters increase the stratification and lead to the shoaling the ML in this region. In the southern part of the NRS (toward the CRS), the MLD distribution is affected by a semi-permanent anticyclone simulated at approximately 25°N, evident as a noticeable deepening of the local ML (Fig. 10a and b), with maximum MLs exceeding 150 m (Fig. 10b). At this location, the local MLD also shows low correlation with atmospheric forcing (Fig. 12), suggesting that MLD deepening is mostly related to the existence of the anticyclone.

In spring the MLD generally shoals to depths less than 30 m, except in the western part of the NRS and in the vicinity of the straits connecting to the two gulfs in the north. Mean MLD in these regions reach about 50 m, while maximum depths may exceed 100 m. Correlation of MLD in the vicinity of the straits of the two gulfs in the north is higher with buoyancy fluxes than with wind stress (Fig. 10b), suggesting that ML deepening is sustained by evaporative heat fluxes and convective mixing. In summer, MLs are generally shallow, remaining less than 30 m. Relatively deeper MLDs are sustained in its western parts, where correlation with wind stress is higher, and MLD may exceed 50 m along the western coasts (Fig. 10b).

*Central Red Sea.* The deepest MLs in the CRS develop mainly along the axis of the basin and in its eastern parts (Fig. 10); they are shallower than those in the NRS, with monthly means not exceeding 50 m and maximum depths generally less than 120 m. In winter, surface buoyancy loss in this region is weaker than it is further north (Fig. 11c), however it exhibits high correlation with MLD (Fig. 12a). Wind stress shows higher correlation with MLD in its eastern part and increases towards the south. The general northward flow of the upper layers that carries warmer and less saline waters from the south, initially follows a westward path leading to shallower MLDs in the western parts of the CRS, while its influence eventfully shifts toward the northeast.

In summer, MLs in the CRS are generally shallow (mean less than 20 m, reaching a maximum of 40 m, Fig. 10), except in the south (around 18°N), where a strong anticyclone develops in response to the Tokar Jet winds locally deepens the ML to more than 60m depth (Fig. 10b). The anticyclone is part of a dipole eddy system, with a cyclone in the north having a weaker but noticeable influence on the MLD (Sofianos and Johns, 2015). MLD is strongly correlated with wind stress over the Tokar anticyclone, whereas its effects extend also to the eastern part of the CRS (Fig. 12b). The correlation of MLD with buoyancy fluxes is weak over this region (Fig. 12a). In contrast, correlation of MLD is higher with buoyancy forcing in the region where the cyclonic eddy persists (Fig. 12a). Despite the cease of the Tokar Jet during early September (Jiang et al., 2009), the semi-permanent dipole eddy system persists longer, affecting the MLs until October (Fig. S7).

*South Red Sea.* Generally, the SRS has relatively shallower MLs than the rest of the basin (Fig 10e). During summer, MLs are restricted to a thin upper layer (less than 20 m), relatively deeper in its western parts (locally reaching 30 m depth). This is a result of the intense surface heat gain and the southward transport of warmer and saline surface waters from the CRS that both increase the stratification. The depth of this upper layer is also bounded from below by the intrusion of the fresher and cooler GAIW. The SRS is affected by strong southward winds during this period, which show a higher correlation with the MLD throughout the region (Fig. 12b).

Following the monsoon reversal in September, warmer and fresher surface waters from the GoAd enter the RS through the BAM. They mainly affect the western parts of the SRS (Yao et al., 2014b), increasing the upper layer stratification and shoaling the MLDs (Fig. 10). The deepest MLs (up to 80 m) develop in the eastern part of the SRS and along the meridional axis of the basin in winter (Fig 10b), where the MLD shows higher correlation with the southern winds entering the RS through the BAM (Fig. 12b). Buoyancy forcing shows low correlation (Fig. 12a), as the surface heat loss in SRS is weak during winter (Fig. 8). The zonal extent of deep MLs is restricted to the central part of the SRS by the presence of the extended coastal shallow areas on both sides of the region (Fig. 10). This increase in MLD, particularly in the vicinity of the BAM (Fig. 10), lasts until April (Fig. 11e). In May, several factors contribute to ML shoaling: these include the gradual weakening of the southern winds, the increase in surface heat gain, and the reestablishment of the southward transport of warmer surface waters from inside the RS.

## 6. Summary and discussion

This study provided a detailed description of the seasonal and spatial evolution of the MLs in the RS, based on a 15-year numerical simulation. The model was implemented with a higher resolution (1/100°, 50 vertical layers) than previous models of the RS and forced by a recently developed high-resolution (approximately 5 km) regional atmospheric reanalysis product, shown to successfully reproduce the spatial and temporal variability of the atmospheric fields. The model outputs were extensively validated against available observations, showing that it can reproduce the seasonal evolution of the upper layer temperature and salinity distributions, the main features of the RS circulation dynamics and their spatiotemporal variability.

Several different methods, including both property difference-based criteria (using density and temperature thresholds), and gradient-based criteria, were evaluated to determine the MLD from both the CTD profiles and the model outputs. The most robust MLD estimates were obtained using the relative variance method of Huang et al. (2018) applied on potential density. The method performed well for different vertical structure's characteristics, as it does not depend on subjective thresholds tailored to specific thermohaline conditions, and showed less sensitivity to noisy data. The modelled MLD estimates were generally in good agreement with those derived from the CTD observations during both seasons and throughout the basin. Differences in the MLD estimates were generally small throughout the basin, with RMSD of 12.6 m during summer and 15.6 m during winter.

The seasonal variability of the MLD was then analyzed based on the model-based estimates. The simulation revealed a strong spatiotemporal variability that emphasize the difficulty of studying the MLDs in the RS by combining the sparse observations from different locations and in different years. In summer, the MLs are generally shallow almost everywhere in the RS (seasonal means approximately 20 m deep), with the exception of a local deepening in the southern part of the CRS related to the Tokar anticyclone, where MLDs occasionally exceed 60 m depth. Relatively deeper MLs are also sustained in the vicinity of the Straits of Tiran inside the GoAq and in the western part of the NRS (exceeding 50 m locally), and in the western parts of SRS (reaching 50 m). In winter, the MLs are deepest in the northern parts of the basin, and generally shoal toward the south. MLs deepening starts in early autumn throughout the RS and deep MLs are sustained until the end of March, while the deepest MLs occur in January and February, especially in the western parts of the NRS and in the GoAq. Monthly mean MLD over these regions exceed 180 m, while locally maximum MLD reaches more than 250 m in the NRS and exceed 400 m in the GoAq. However, large standard deviations in the climatological monthly mean values throughout the basin indicate the development of significantly deeper MLs in some years.

Analysis of the relative contribution of atmospheric forcing components in inducing vertical mixing revealed that the seasonal variability is mostly controlled by the variability of the surface buoyancy fluxes. They also show higher correlation with the seasonal MLD evolution. The strongest buoyant mixing is in the northern Red Sea and GoAq in winter, associated with the deepest MLs. Buoyancy forcing is driven predominantly by its heat flux component. Although freshwater fluxes (driven primarily by evaporation) in the RS region dominate the annually averaged surface buoyancy forcing, its seasonal variability is too weak to significantly affect the seasonal variability of the total buoyancy flux. Due to the cumulative effect of wintertime surface buoyancy loss, the deepest MLs in the RS develop after the peak in the surface heat loss.

Wind induced mixing has a higher contribution in summer, when buoyancy forcing acts to increase the stratification. During autumn, the wind contribution gradually decreases, as buoyancy loss starts to increase everywhere in the RS. The wind driven mixing remains relatively strong only in the SRS due to the monsoon-driven southern winds that are channeled through the BAM. Its influence becomes less significant in winter when convective processes dominate vertical mixing. In early spring, restratification causes the MLs to shoal, whereas relatively deep MLs in the northern parts of the basin (NRS and the GoAq) and in the SRS near the BAM also correlate strongly with winds. Channeling of winds through the straits and jets blowing through mountain gaps induce strong mixing and locally also increase buoyancy loss that drive local MLD deepening through convective mixing. Wind-induced mixing dominates over convective processes in the northern parts of the basin, due to channeling of the northern winds through the narrow Gulfs that exit towards the NRS.

Despite the generally high correlation of the simulated MLD with the atmospheric forcing components, its temporal and spatial variability suggests that advective fluxes also influence its development. For example, during spring the correlation with buoyancy fluxes is weak in the CRS and NRS, despite the onset of restratification. MLD is also influenced by the lateral advection of heat and salt, evident throughout the basin and especially in the vicinity of straits, such as in the SRS near the BAM strait and in the south GoAq. Finally, semi-permanent eddies and gyres show a strong influence on MLD by locally displacing the isopycnals, creating troughs and ridges in the MLD distribution.

The strong spatiotemporal variability of atmospheric buoyancy and momentum forcing, thermohaline circulation and mesoscale activity are all imprinted on the MLD distribution, suggesting an interplay between the surface air–sea exchanges and internal processes governing the MLD variability. While this study mainly examined the relation of air–sea fluxes with the seasonal evolution of the MLs, the quantification of the effects of internal processes will be specifically addressed in our next study.

**Acknowledgments**

The research reported in this publication was supported by King Abdullah University of Science and Technology (KAUST) under the Competitive Research Grant Program (CRG), Grant #UFR/1/2979-01-01. Model simulations and postprocessing of its outputs were performed on the KAUST supercomputing facility Shaheen-II, Saudi Arabia. We are grateful to the KAUST HPC team for their valuable support during the generation and processing of the model outputs. We also thank the UK MetOffice Ocean for providing the Operational SST and Sea Ice Analysis (OSTIA) products. Data used in this study will be made publicly available in compliance with the FAIR (Findability, Accessibility, Interoperability, and Reusability) Data standards. OSTIA data are available from the Copernicus Marine Environmental Monitoring Service (CMEMS) web portal (http://marine.copernicus.eu/).

# Supporting Information for

# Seasonal Evolution of Mixed Layers in the Red Sea and the Relative Contribution of Atmospheric Buoyancy and Momentum Forcing

## S1. Remotely sensed data used for model validation

Daily averaged SST is validated against corresponding fields from the "Operational SST and Sea Ice Analysis" (OSTIA), provided by the UK Met Office (Donlon et al., 2012), on a 0.05° horizontal resolution grid, available from 1985 to present. This dataset is based on a blend of microwave and infrared satellite measurements, a product developed by the Group for High Resolution SST (GHRSST). The OSTIA product has been validated by inter-comparisons with other historical datasets (e.g. Reynolds Optimal Interpolation and Hadley Centre SST), and is continuously monitored and validated with in-situ measurements (Donlon et al., 2012). This is the highest resolution remotely sensed dataset available for the Red Sea that covers the entire period of our model simulation (2001-2015). OSTIA SST has been previously used to analyze the RS SST variability and trends, and has been successfully validated against other available satellite and historical reconstructions in this region (Genevier et al., 2019; Krokos et al., 2019).

Sea level anomaly (SLA) derived from satellite altimeter data is used to evaluate the persistent circulation features and mesoscale activity in the Red Sea. SLA is defined as a measure of the sea surface height with respect to a mean sea level, computed by long term averaging altimeter data. The dataset used in this study is the latest version of the SSALTO/DUACS products based on multi-mission satellite altimeter observations provided by AVISO (ftp://ftp.aviso.oceanobs.com/global/dt/upd/msla/merged/). The product utilizes information from a multitude of satellite altimetry data that ensure a better spatial coverage and fewer mapping errors than single satellite products (Ducet et al., 2000). The product is provided on a 0.25° grid with a monthly resolution for the period 1993-2015.

## S2. Comparison with remotely sensed SST

A comparison of modeled and satellite-derived daily SST fields (OSTIA), over the time period of the model simulation, averaged over both winter (October-April) and summer (May-September) period is shown in Fig. S1. The model SST generally agrees better with the satellite-derived SST during the winter period, and differences are larger during the summer period. Averaged over the entire RS, the Root Mean Squared Difference (RMSD) between the satellite and modeled SST is 0.76 ± 0.23 °C during the winter and 0.91 ± 0.36 °C during the summer periods. The correlation coefficients between the daily averaged SST fields are 0.97 ± 0.02 and 0.93 ± 0.05, for winter and summer periods respectively. The RMSD is less than 0.8°C almost everywhere in the RS (except in the shallowest areas, Fig. S1a). Overall, the correlation coefficient is spatially uniform, exceeding 0.95 almost everywhere in the RS (Fig. S1b).

During the winter period, the agreement between the model and satellite SST is good almost everywhere in the basin. Some mismatch is apparent in the CRS and the adjacent parts of the NRS, likely due to the existence of short-lived eddies prevalent in the region. A good agreement between the two SST datasets is observed in the NRS and the two northern gulfs during winter (RMSD of approximately 0.5°C). This is particularly important, as these regions present the highest seasonal variability and the deepest MLs.

The largest differences occur in the shallow coastal areas of the SRS, especially during the summer period. This is expected, as these regions present a challenge for both satellite remote sensing and model simulation. Extended shallow regions in the SRS are relatively isolated from the open waters, due to the existence of complex coral reefs structures, and exhibit extreme variability in temperature. Accurate simulation of the local circulation is hindered by uncertainties in the bathymetry. Additional difficulties arise from high concentrations of atmospheric aerosols (especially dust) and cloud coverage in these regions (Dreano et al., 2016; Kumar et al., 2018), which impact the accuracy of both satellite observations (Donlon et al., 2012) and atmospheric modelling products (Viswanadhapalli et al., 2017). Remotely sensed data are also known to have large uncertainties in these areas due to shallow depths and land contamination by the large number of scattered islands (Donlon et al., 2012; Stark et al., 2007). Despite these, the RMSD between the model and satellite SST datasets is generally less than 1°C, except for a narrow band in the near coastal region of the SRS during the summer period.

## S3 Comparison with remotely sensed Sea Level Anomalies

Climatological averages of the mean SLA from satellite observations and the model simulation are presented in Figure S2. The dominant circulation features are reflected in the spatiotemporal variability of the SLA (Zhan et al., 2014). Observed SLA features represent the seasonal cycle of the general upper layer circulation and the impact of persistent cyclonic and anticyclonic eddies. The spatial and seasonal distribution of SLA in the model shows almost identical patterns and similar intensity with the satellite derived data, suggesting a qualitative consistency in the representation of circulation patterns. SLA reveals a seasonaly varying mean sea levels with alternating meridional gradients, which are generally stronger in the model simulation. This is expected as satellite altimetry cannot resolve small scale currrents and is known to underestimate the boundary currents near the coasts that are especially strong in the RS (Bower and Farrar, 2015).

A succession of cyclones and anticyclones is evident throughout the year, with seasonal variations in their distribution. The main features during summer are the two strong anticyclones at 18 ºN and 23 ºN, and a series of smaller cyclonic features in the central and north parts of the basin. A west-east gradient in the south has been suggested to occur due to the strong northern winds, induced by the southward shift of the Tokar jet winds over the SRS during summer (Sofianos and Johns, 2003). During Autumn, the surface circulation intensifies towards the north throughout the basin and strong eddy activity appears in the CRS, as well as a strengthening of the cyclonic circulation in the NRS. During winter, SLAs exhibit higher spatial variability and the west-east gradient reverses, with higher sea levels in the eastern part. The meridional gradient during winter has been linked to the seasonal southern winds in the south, and to the northward eastern boundary currents in the north (Sofianos and Johns, 2003; Bower and Farrar, 2015).

## S4 Validation of the model circulation

This section focuses on the representation of the general and mesoscale circulation of the model simulation. We first evaluate the water exchange between the RS and the Gulf of Aden, as representative of the thermohaline-driven overturning circulation. We then perform a comparison with available current velocity observations to evaluate the representation of the energetic mesoscale field.

### S4.1. Exchange flows through the Bab-Al-Mandeb Strait

The exchanges through the BAM provide a benchmark for testing the performance of general circulation models of the RS (Bower and Farrar, 2015; Sofianos and Johns, 2015). The volume transport from the model for each characteristic water mass has been computed following the method of Xie et al. (2019). A three-layer pattern (with the surface outflow) was fitted into the model output in summer, and a two-layer pattern (with the surface inflow) was fitted in winter. The individual layers were then defined by their transport direction, and transport estimates were computed as a depth-integral over such defined layers (Fig. S1). During winter, the upper layer inflow (GASW) through the BAM is balanced by a deeper outflow (RSOW), whilst during summer, the three-layer pattern consists of a surface (RSSW) and a deep (RSOW) outflow, balanced by an inflow at intermediate depths (GAIW). Since the upper layers (GASW and RSSW) can be distinguished by the sign of their volume transport, for simplicity, both are presented and denoted as surface waters (SW). A summary of the volume transport estimates from observations at the BAM, compiled by Sofianos and Johns (2015), is used hereafter for the comparison with the model results.

The timing of the seasonal evolution of volume transport agrees well with estimates based on direct measurements (Murray, 1997), and previous modeling studies (Yao et al., 2014a; Xie et al., 2019). Transition from the winter (three-layer) to the summer (two-layer) exchange pattern takes place in early June and lasts till late September (Fig. S1). During the winter period, RSOW and the corresponding SW flow rates each exceed 0.5 Sv, reaching a maximum value of 0.6 Sv in February. During summer, the RSOW gradually weakens, and the GAIW intrusion is mainly balanced by the SW outflow. The RSOW outflow reaches a minimum in August, with an estimate of 0.05 Sv, and during the same period, the volume transport of the GAIW intrusion reaches a peak value of 0.3 Sv (Fig. S1), both closely matching the values reported from observations (Fig. 8 in Sofianos and Johns, 2015). Finally, the model annual mean transport of RSOW (0.35 Sv, Fig. S1) is also nearly identical to the reported transport of 0.36 Sv in Sofianos and Johns (2015).

### S4.2. Comparison with Current observations

Regional observations and numerical simulations of the RS have revealed a complex mesoscale circulation, with energetic, quasi-stationary eddies and gyres and narrow boundary currents, while transports associated with these features have been estimated to an order of magnitude larger than the overturning circulation. (e.g. Quadfasel and Baudner, 1993; Sofianos and Johns, 2003; Yao et al., 2014a,b).

Few observational datasets of current velocities that allow the spatial and temporal verification of the model outputs are available (Clifford et al. 1997). Two hydrographic and current surveys of the eastern Red Sea were carried out on the R/V Aegaeo in March 2010 and September–October 2011 (Bower and Farrar, 2015). The surveys provided continuous measurements of current velocities from a vessel-mounted Acoustic Doppler Current Profiler (ADCP) down to about 600 m depth along the entire cruise track. As the dataset was not available for this study, a comparison of the major circulation features in the upper layers is here based on results presented in a recent study by Bower and Farrar (2015), covering the eastern half of the Red Sea north of 22°N. The schematic diagrams of the upper half of the RS (Fig. 9, in Bower and Farrar, 2015), illustrating the major circulation features and the corresponding daily model outputs during the period of the surveys, are provided in Figure S4. Observations reveal complex patterns of eddies and meandering currents throughout the cruise track, associated with anticyclonic eddies, with diameter of the order of 100-200 km. The model reproduces the main features of the upper layer circulation and the spatial variability of potential temperatures, with a generally good agreement in the positions and scales of the observed eddies. The simulation also captures the seasonal difference in the dominant circulation features, especially the intensification and meridional spread of the cyclonic gyre in the NRS.

Observations of current velocities along the RS were collected during an expedition aboard the R/V Maurice Ewing from 4 August to 19 August 2001 (Sofianos and Johns, 2007), which captured the characteristics of the summer RS circulation. The expedition followed an along-axis transect of the basin and current velocities were measured using a vessel-mounted ADCP. The observed current velocities, overlayed on the model velocity field are presented for 4 different depths (Fig. S5, from left to right: 14, 46, 78 and 126 m). Overall, the model captures all the main circulation features and their relative intensity throughout the RS, as well as their variability with depth.

In the SRS the velocity field exhibits a high spatial variability in both the model and observations. The surface outflow appears to strengthen from west to east, as waters flow towards the strait. The flow direction reverses at the intermediate depths, as a result of the intermediate water intrusion, which is confined in the deep along-axis tranch of the SRS (Fig S5. 46 and 78 m). In deeper layers (Fig S5, 146 m), a very weak flow towards the south is observed.

Northward, the CRS measurements reveal an intense anticyclone accompanied by a smaller cyclone in its north. This feature is associated with the quasi-stationary dipolar eddy in response to the Tokar Jet winds (Bower et al., 2013; Zhai and Bower, 2013; Zhan et al., 2018). The first of the eddies, a strong anticyclone, presents the highest velocities captured throughout the expedition and are the highest simulated throughout the RS (~0.6 m/s, in both the observations and the model). Its influence gradually weakens with depth, with a very weak signal compared to the surface velocities remaining at 146 m depth. A weaker cyclonic feature in its north is also well resolved by the simulation, however with a small southward shift and a slight underestimation of its intensity. Both observations and the simulation show a maximum depth of its influence at about 80 m.

Following the along-axis survey, observed velocities reveal a anticyclonic circulation located at 23°N, also observed during the KAUST-WHOI cruise in autumn of 2011 (Bower and Farrar, 2015). Both its scale and intensity along depths is realistically represented in the simulation. This eddy is one of the most characteristic semi-permanent features in the NRS that persists for the most part of the year (Clifford et al., 1997; Sofianos and Johns, 2003). Its existence has been attributed to the seasonal buoyancy driven basin-scale circulation rather than wind forcing (Chen et al., 2014). It has influence in depth, which is well simulated, remaining strong in the deepest layer evaluated (Fig S5, 146 m).

In the NRS, both datasets show a reduced velocity field, organized in a cyclonic gyre centered between 26ºN and 27ºN. On its western part, the observations shows a southward intensification of the flow parallel to the coast also evident in the simulation. Finally, velocities in the GoS, show a southeastward surface flow and an nothwestward flow bellow. Velocities are less pronounced in Figure S5, mainly due to the compromise in the relative magnitude with the stronger velocity field in the RS.

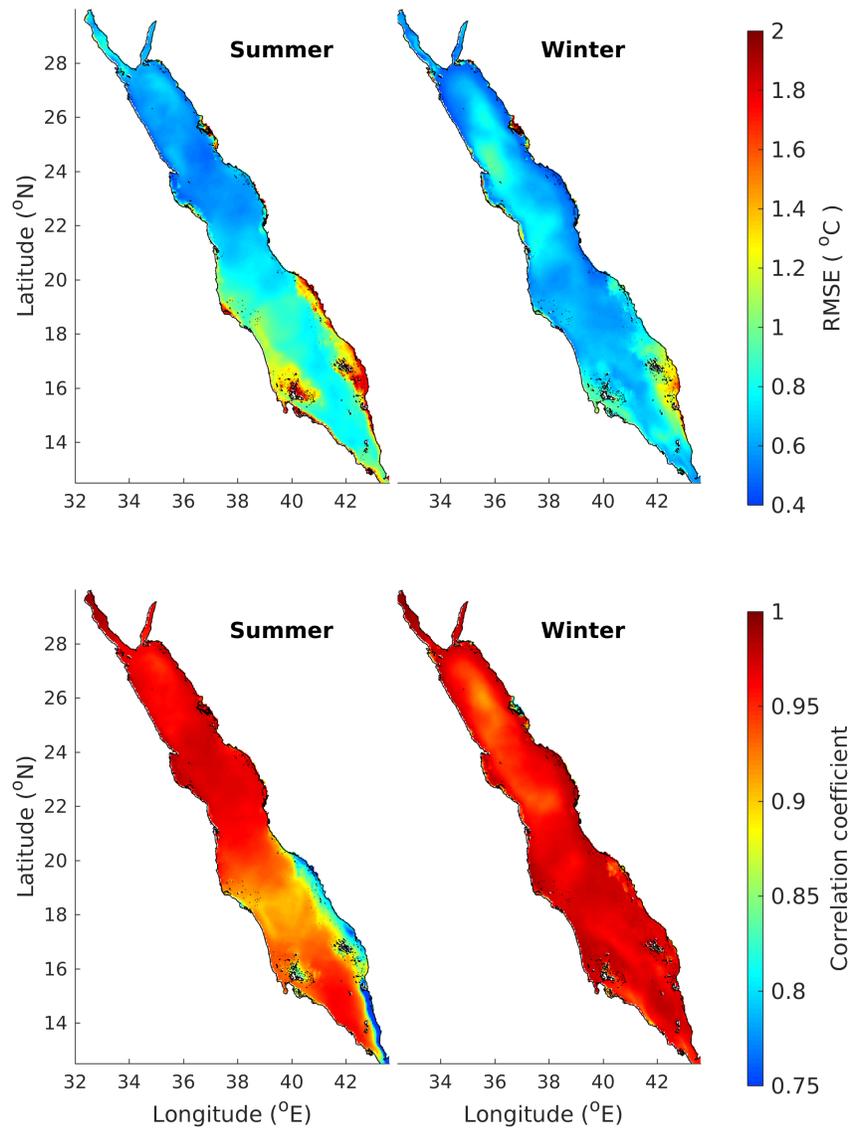

Figure S1. Upper panel: The Root Mean Square Difference (RMSD) between the daily modeled and satellite derived SST (OSTIA), time averaged over the model simulation period (2001-2015), separately for the winter (October-April) and summer (May-September) periods. Lower panel: The correlation coefficient between the two SST estimates.

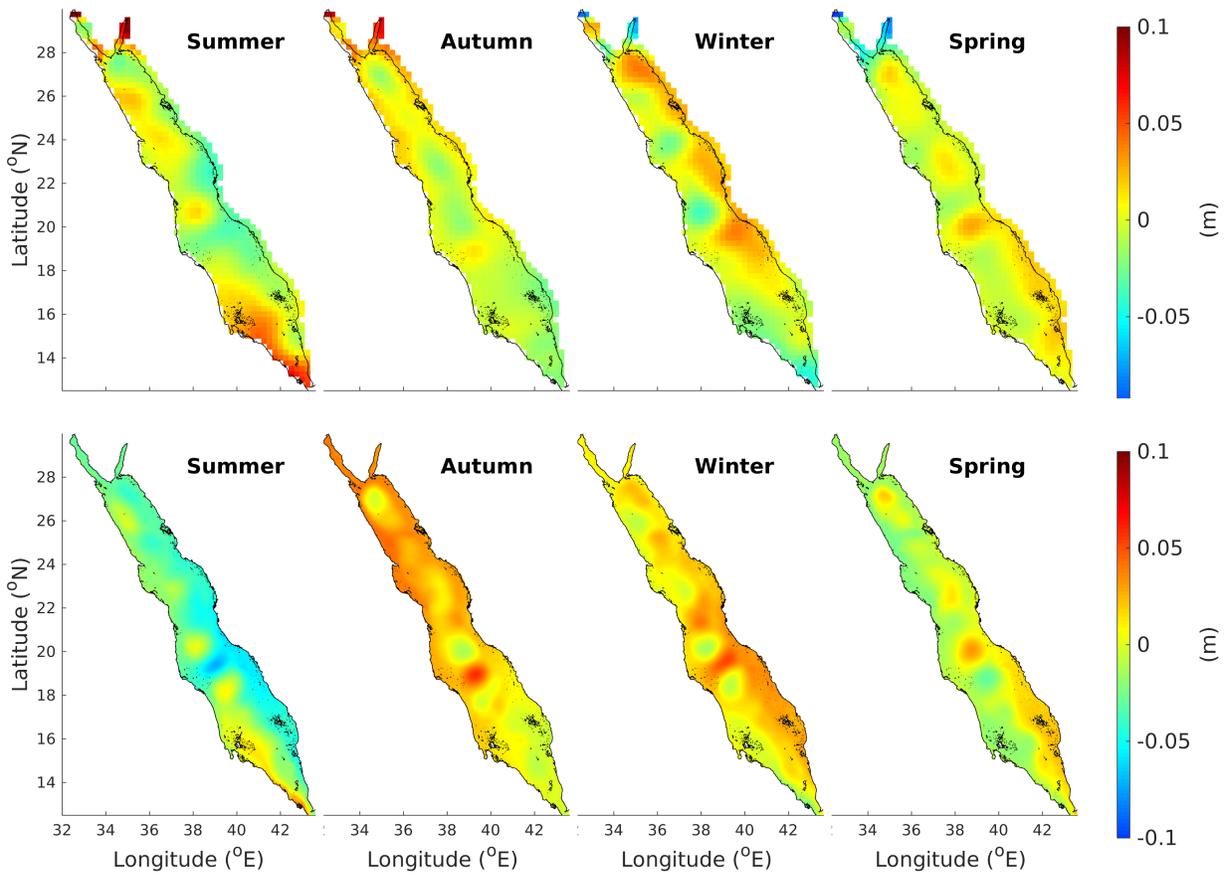

Figure S2. Upper panel: Seasonal climatology of sea level anomaly derived from satellite altimetry data (AVISO) for the period (2001-2015), in: summer (June-August), autumn (Sep-Nov), winter (Dec- Feb) and spring (March-May). Lower panel: The same, except from the model output.

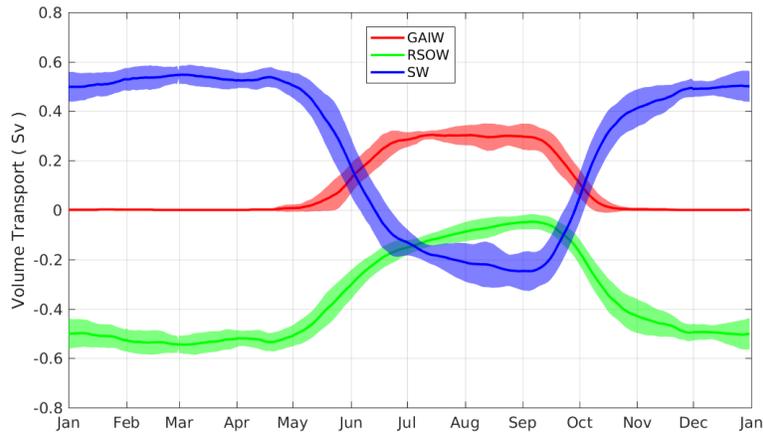

Figure S3. Climatology of the volume transport at the strait of BAM of the different water masses SW, GAIW and RSOW (lines), and their respective standard deviations (shaded areas) for the period 2001-2015 from daily mean model outputs. Daily mean model outputs have been filtered with a 30-day low pass filter to emphasize the seasonal cycle.

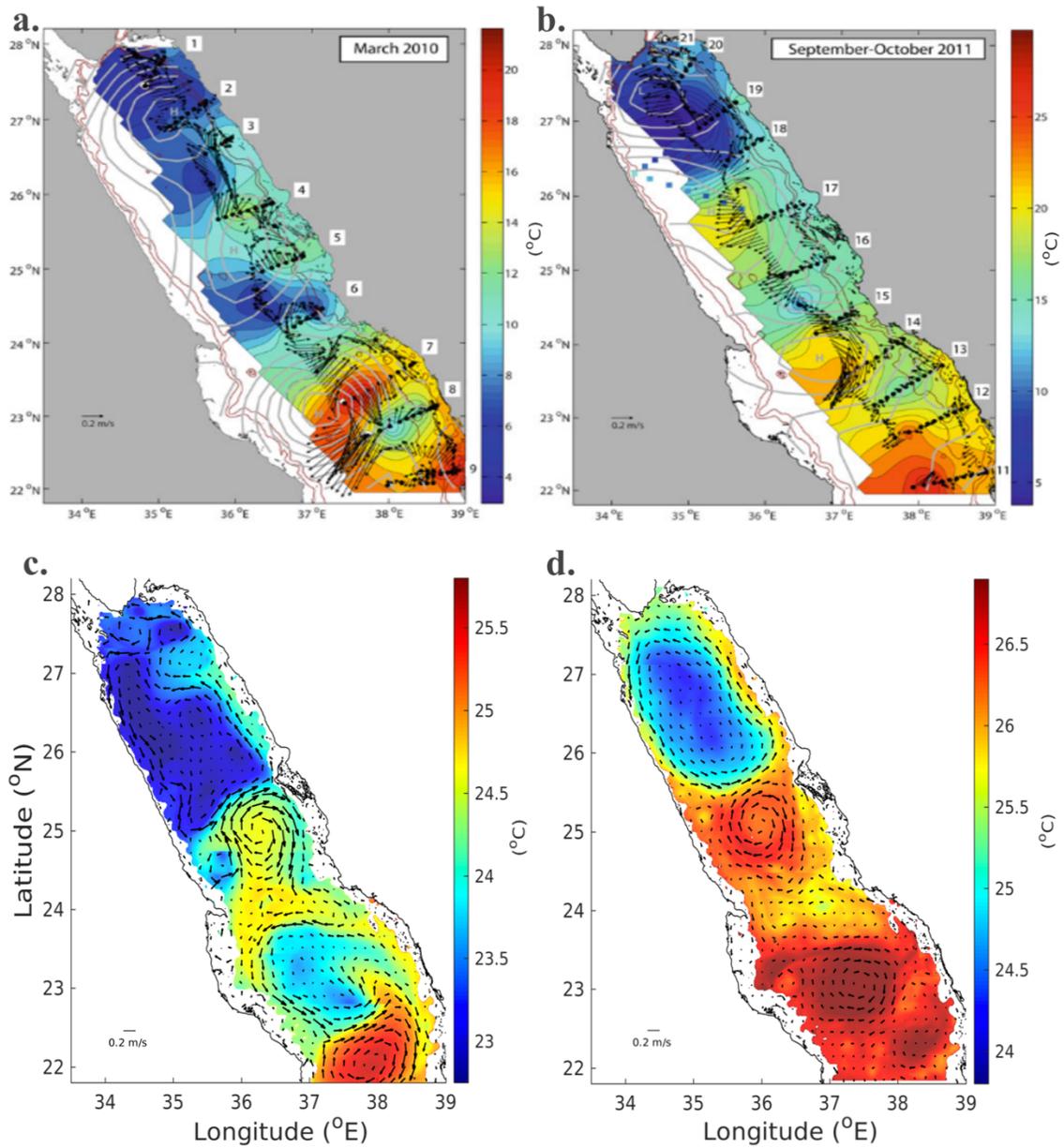

Figure S4. (a, b) Schematic diagrams illustrating major circulation features during two surveys carried out on the R/V Aegaeo in March 2010 and September–October 2011, and drifter observations, superimposed on the 0–200 m vertically averaged potential temperature (color shading) (adopted from fig. 9 in Bower and Farrar (2015)). (c, d) The corresponding 0–200 m vertically averaged current velocities (black arrows) and potential temperature (color shading, values shown for depths greater than 200m) from daily model outputs produced in this study, averaged during the two surveys' periods.

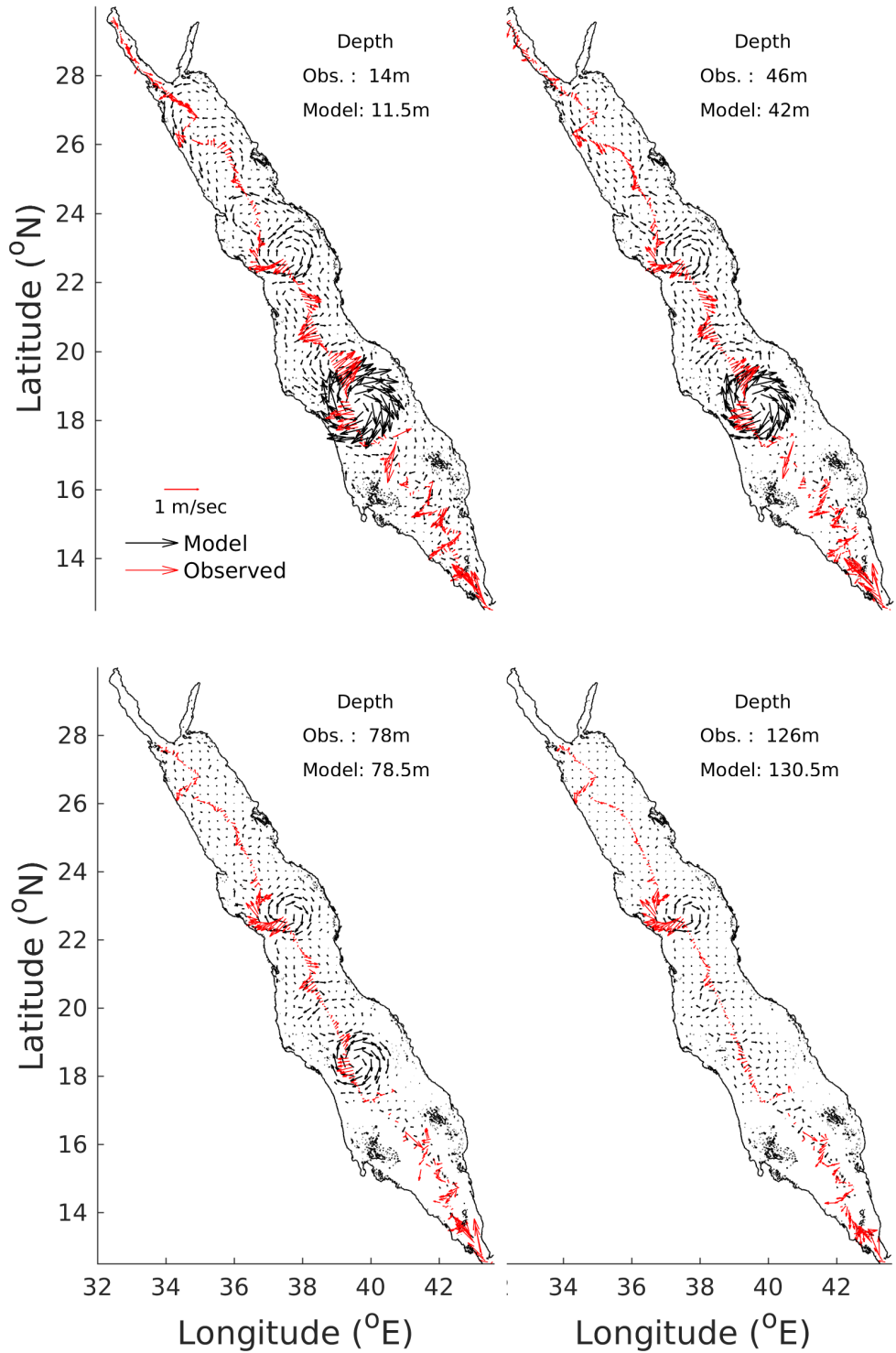

Figure S5. Velocity profiles at 4 different depths (from left to right: 14, 46, 78 and 126 m) sourced from current profiler observations (red arrows) taken along-track the R/V Meteor expedition

during August 2001, and the corresponding simulated velocity profiles from daily model outputs (black arrows).

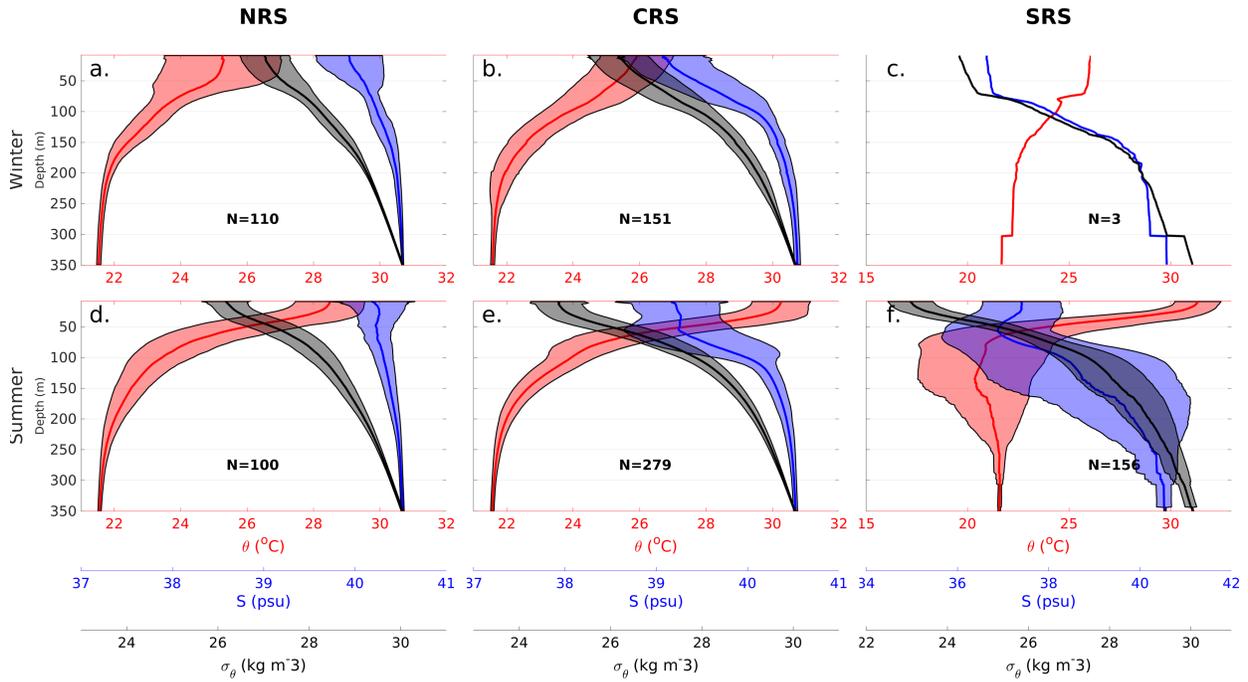

Figure S6. Mean upper layer (top 350m) vertical profiles of salinity S, potential temperature $\theta$, and potential density $\sigma_\theta$, obtained from the CTD observations listed in Table 1, over the (a,b) NRS, (c,d) CRS and (e,f) SRS regions, in (a,c,e) winter and (b,d,f) summer. Shading shows the corresponding standard deviation and N denotes the number of observations considered.

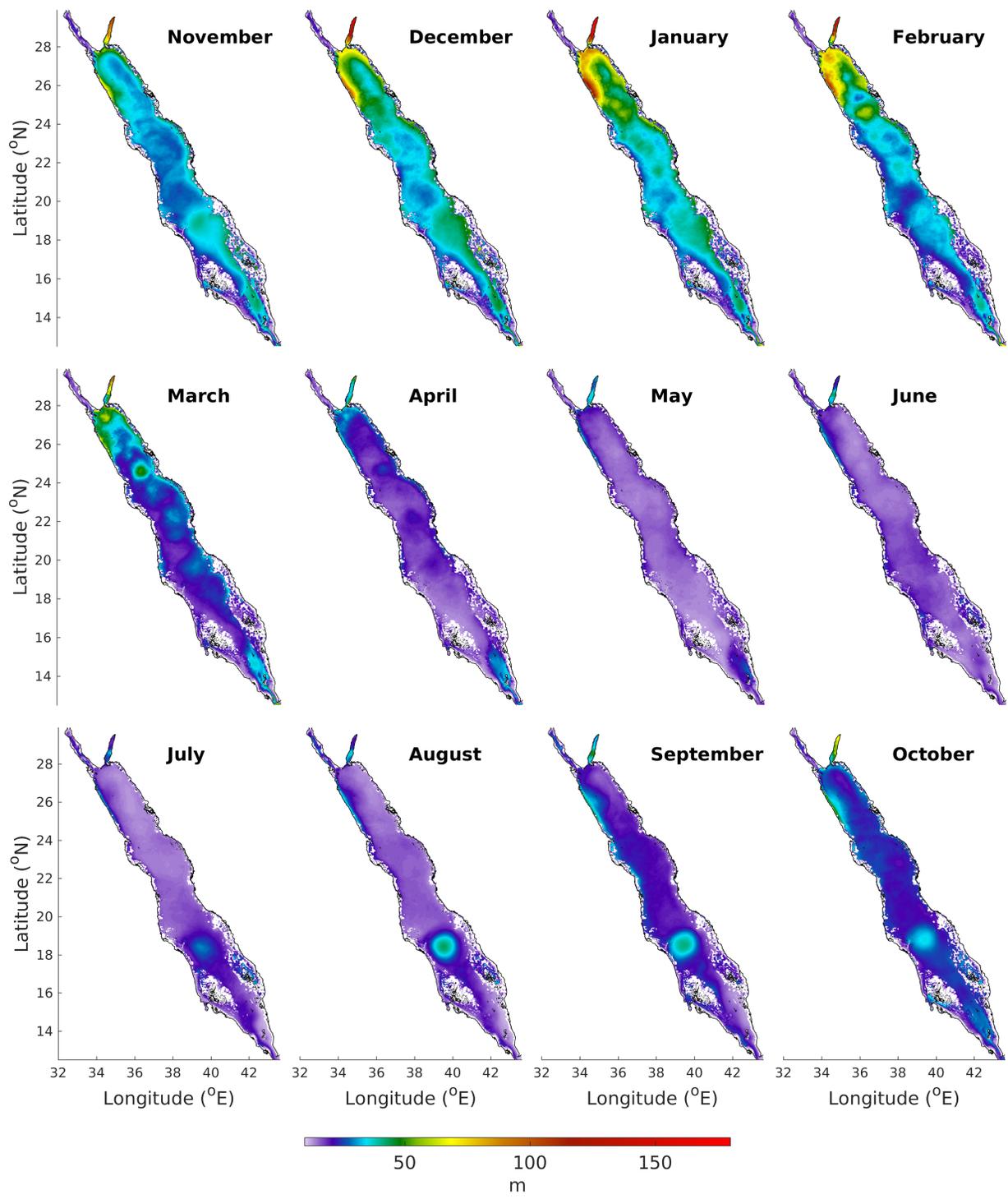

Figure S7. Monthly climatology of the MLD estimates, obtained using the method of Huang et al. (2018) on daily model outputs, for the model simulation period (2001-2015).